\documentclass[aps,prd,onecolumn]{revtex4}

\usepackage{graphicx}  
\usepackage{amsmath}
 
 \usepackage{amssymb}
 
\newcommand{\cc}{cosmological constant}
\newcommand{\del}{\partial}
\newcommand{\dphi}{\partial_i \phi \partial^i \phi}

  \def\rb{\right)}
\def\lb{\left(}
\def\frab#1#2{\left({#1\over#2}\right)}
\def\fra#1#2{{#1\over#2}}

\def\bld#1{{\bf #1 }}
\def\pdov#1#2{{\partial#1\over \partial#2}}

\def\eq#1{{Eq.~(\ref{#1})}}
\def\eqs#1{{Eqs.~(\ref{#1})}}
\def\fig#1{{Fig.~\ref{#1}}}
\def\figs#1{{Figs.~\ref{#1}}}

\def\pdov#1#2{\frac{\partial#1}{ \partial#2}}

\def\rb{\right)}
\def\lb{\left(}       
\def\frab#1#2{\left(\frac{#1}{#2}\right)}         
\def\fra#1#2{\frac{#1}{#2}}
\def\bld#1{{\bf #1 }}

\newcommand{\de}{{\rm d}}

\newcommand{\alp}{\alpha}

\newcommand{\xeta}{{\bf x}, \eta}

\begin{document}

\title{ADVANCED TOPICS IN COSMOLOGY: A PEDAGOGICAL INTRODUCTION}

\author{T.~Padmanabhan}
\affiliation{IUCAA,
Post Bag 4, Ganeshkhind, Pune - 411 007\\
email: nabhan@iucaa.ernet.in}


\begin{abstract}
These lecture notes provide a concise, rapid and pedagogical introduction to several
advanced topics in contemporary cosmology. The discussion of thermal history of the 
universe, linear perturbation theory,  theory of CMBR temperature anisotropies and
the inflationary generation of perturbation are presented in a manner accessible
to someone who has done a first course in cosmology. The discussion of dark energy
is more research oriented and reflects the personal bias of the author. Contents:
(I) The cosmological paradigm and Friedmann model; (II) Thermal history of the 
universe; (III) Structure formation and linear perturbation theories; (IV) Perturbations 
in dark matter and radiation; (V) Transfer function for matter perturbations; (VI) Temperature anisotropies of CMBR; (VII) Generation of initial perturbations from inflation; (VIII) The dark energy.
\end{abstract}

\maketitle 

\tableofcontents
 
 \section{The Cosmological Paradigm and Friedmann model}

Observations show that the universe is fairly homogeneous and isotropic at scales
larger than about $150h^{-1}$ Mpc where 1 Mpc $\simeq 3 \times 10^{24}$ cm and $h\approx 0.7$ is a parameter related to the expansion rate of the universe. The conventional --- and highly successful --- approach to cosmology
separates the study of large scale ($l\gtrsim 150h^{-1}$ Mpc) dynamics of the universe from the issue of structure formation at smaller scales. The former is modeled by  a homogeneous and isotropic distribution of energy density; the latter issue is
addressed  in terms of gravitational instability which will amplify the small perturbations in the energy density, leading to the formation of structures like galaxies.
In such an approach, the expansion of the background universe is described by the metric (We shall use units with with $c=1$ throughout, unless otherwise specified):
\begin{equation}
   ds^2 \equiv dt^2 -a^2d\textbf{x}^2
  \equiv dt^2 - a^2(t) \left[ d\chi^2 + S_k^2(\chi) \left( d\theta^2 + \sin^2 \theta d\phi^2 \right) \right]
  \label{frwmetric}
  \end{equation}
 with $S_k(\chi) = (\sin \chi, \chi, \sinh \chi) $ for the three values of the label $k= (1,0,-1)$.
The function  $a(t)$
 is governed by the equations: 
\begin{equation} 
\frac{\dot a^2+k}{a^2} =\frac{8\pi G\rho}{3};\qquad d(\rho a^3)=-pda^3
\label{frw}
\end{equation}
The first one relates expansion rate of the universe to the energy density $\rho$ and $k=0,\pm 1$ is a parameter which characterizes the spatial curvature of the universe. 
 The second equation, when coupled with the equation of state 
$p=p(\rho)$ which relates the pressure $p$ to the energy density, determines the evolution of energy density  $\rho=\rho(a)$ in terms of the expansion factor of the universe.
 In particular if $p=w\rho$ with (at least, approximately) constant $w$ then, $ \rho \propto a^{-3(1+w)}$ and (if we  further assume $k=0$, which is strongly favoured by observations) the first equation in Eq.(\ref{frw}) gives
$ a \propto t^{2/[3(1+w)]}$. We will also often use the redshift $z(t)$, defined as $(1+z)=a_0/a(t)$ where the subscript zero
denotes quantities evaluated at the present moment. in a $k=0$ universe, we can set $a_0=1$ by rescaling the spatial coordinates.
 
 It is convenient to measure
the energy densities of different components in terms of a \textit{critical energy density} ($\rho_c$) required to make $k=0$ at the present epoch. (Of course, since $k$ is a constant,
it will remain zero at all epochs if it is zero at any given moment of time.) From Eq.(\ref{frw}), it is clear that $\rho_c=3H^2_0/8\pi G$ where $H_0\equiv (\dot a/a)_0$ --- called the Hubble constant ---
is the rate of expansion of the universe at present. 
 Numerically
  \begin{eqnarray}
 \rho_c  = \frac{3H_0^2}{8\pi G} & =& 1.88 h^2 \times 10^{-29}\ {\rm gm\ cm}^{-3}= 2.8 \times 10^{11} h^2 M_\odot \ {\rm Mpc}^{-3}\nonumber\\
& =& 1.1 \times 10^4 h^2 \ {\rm eV\ cm}^{-3} = 1.1 \times 10^{-5} h^2 \ {\rm protons \ cm}^{-3}
\end{eqnarray} 
 The variables $\Omega_i\equiv \rho_i/\rho_c$ 
will give the fractional contribution of different components of the universe ($i$ denoting baryons, dark matter, radiation, etc.) to the  critical density. Observations then lead to the following results:

(1) Our universe has $0.98\lesssim\Omega_{tot}\lesssim1.08$. The value of $\Omega_{tot}$ can be determined from the angular anisotropy spectrum of the cosmic microwave background radiation (CMBR; see Section \ref{sec:tempcmbr})  and these observations (combined with the reasonable assumption that $h>0.5$) show\cite{cmbr}  that we live in a universe
with critical density, so that $k=0$.

(2) Observations of primordial deuterium produced in big bang nucleosynthesis (which took place when the universe
was about few minutes in age) as well as the CMBR observations show\cite{baryon}  that  the {\it total} amount of baryons in the
universe contributes about $\Omega_B=(0.024\pm 0.0012)h^{-2}$. Given the independent observations\cite{h} which fix $h=0.72\pm 0.07$, we conclude that   $\Omega_B\cong 0.04-0.06$. These observations take into account all baryons which exist in the universe today irrespective of whether they are luminous or not. \textit{Combined with previous item we conclude that
most of the universe is non-baryonic.}

(3) Host of observations related to large scale structure and dynamics (rotation curves of galaxies, estimate of cluster masses, gravitational lensing, galaxy surveys ..) all suggest\cite{dm} that the universe is populated by a non-luminous component of matter (dark matter; DM hereafter) made of weakly interacting massive particles which \textit{does} cluster at galactic scales. This component contributes about $\Omega_{DM}\cong 0.20-0.35$ and has the simple equation of state $p_{DM}\approx 0$.  The second equation in Eq.(\ref{frw}), then gives $\rho_{DM}\propto a^{-3}$  as  the universe expands which arises from the evolution of number density of particles: $\rho=nmc^2
\propto n\propto a^{-3}.$

(4) Combining the last observation with the first we conclude that there must be (at least) one more component 
to the energy density of the universe contributing about 70\% of critical density. Early analysis of several observations\cite{earlyde} indicated that this component is unclustered and has negative pressure. This is confirmed dramatically by the supernova observations (see Ref.~\cite{sn}; for a critical look at the current data, see Ref.~\cite{tptirthsn1}).  The observations suggest that the missing component has 
$w=p/\rho\lesssim-0.78$
and contributes $\Omega_{DE}\cong 0.60-0.75$. The simplest choice for such \textit{dark energy} with negative pressure is the cosmological constant which is  a term that can be added to Einstein's equations. This term acts like  a fluid with an equation of state $p_{DE}=-\rho_{DE}$; the second equation in Eq.(\ref{frw}), then gives $\rho_{DE}=$ constant as universe expands.

(5) The universe also contains radiation contributing an energy density $\Omega_Rh^2=2.56\times 10^{-5}$ today most of which is due to
photons in the CMBR. 
The equation of state is $p_R=(1/3)\rho_R$; the second equation in Eq.(\ref{frw}), then gives $\rho_R\propto a^{-4}$. Combining it with the result
$\rho_R\propto T^4$ for thermal radiation, it follows that
 $T\propto a^{-1}$.
Radiation is dynamically irrelevant today but since $(\rho_R/\rho_{DM})\propto a^{-1}$ it would have been the dominant component   when
the universe was smaller by a factor larger than $\Omega_{DM}/\Omega_R\simeq 4\times 10^4\Omega_{DM}h^2$.

(6) Taking all the above observations together,  we conclude that our universe has (approximately) $\Omega_{DE}\simeq 0.7,\Omega_{DM}\simeq 0.26,\Omega_B\simeq 0.04,\Omega_R\simeq 5\times 10^{-5}$. All known observations
are consistent with such an --- admittedly weird --- composition for the universe.

Using $\rho_{NR}\propto a^{-3}, \rho_R \propto a^{-4}$ and $\rho_{DE}$=constant   we can  write Eq.(\ref{frw}) in a convenient dimensionless form as
\begin{equation}
{1\over 2} \lb {dq\over d\tau}\rb^2 + V(q) = E 
\label{qomeganr}
\end{equation}
where $\tau = H_0t,a= a_0 q(\tau),\Omega_{\rm
NR} = \Omega_B + \Omega_{\rm DM}$ and 
\begin{equation}
V(q) = - {1\over 2} \left[ {\Omega_R\over q^2} + {\Omega_{\rm
NR}\over q} + \Omega_{DE} q^2\right]; \quad E={1\over 2} \lb 1-\Omega_{\rm tot}\rb.
\label{qveq}
\end{equation} 
This equation has the structure of the first integral for
motion of a particle with energy $E$ in a potential $V(q)$. 
For models with $\Omega  = \Omega_{\rm NR} + \Omega_{DE} =1$,
we can take $E=0$ so that $(dq/d\tau) = \sqrt{V(q)}$. Based on the observed composition of the universe, we can identify three distinct phases in 
the evolution of the universe when the temperature is less than about 100 GeV. At high redshifts (small $q$)
the universe is radiation dominated and $\dot q$ is independent 
of the other cosmological parameters. Then Eq.(\ref{qomeganr}) can be easily integrated
to give $a(t) \propto t^{1/2}$ and the temperature of the universe decreases as
$T\propto t^{-1/2}$. As the universe expands, a time will come
when ($t=t_{\rm eq}$,  $a=a_{\rm eq}$ and $z= z_{\rm eq}$, say)
the matter energy density will be comparable to radiation energy 
density. For the parameters described above, $(1+z_{eq})=\Omega_{NR}/\Omega_R\simeq 4\times 10^4\Omega_{DM}h^2$. At lower redshifts, matter will dominate over radiation and we will
have $a\propto t^{2/3}$ until fairly late when the 
dark energy density will dominate over
non relativistic matter.
This occurs at a redshift of $z_{\rm DE}$ where $(1+ z_{\rm DE}) =( \Omega_{\rm DE}/\Omega_{\rm NR})^{1/3}$.
For $\Omega_{\rm DE} \approx 0.7, \Omega_{\rm NR} \approx 0.3$, this occurs at $z_{\rm DE}\approx 0.33$.
In this phase, the velocity
$\dot q$ changes from being a decreasing function to an increasing function leading to 
an accelerating universe. In addition to these, we believe that the universe probably
went through a rapidly expanding, inflationary, phase very early when $T\approx 10^{14}$ GeV;
we will say more about this in Section \ref{sec:inflation}. (For a textbook description of these and related issues, see e.g. Ref.~\cite{tpsfuv3}.)
 
 Before we conclude this section, we will briefly mention some key aspects of 
  the background cosmology described by a Friedmann model. 
  
 (a) The metric in Eq.(\ref{frwmetric}) can be rewritten using the expansion parameter $a$ or the redshift $z = (a_0/a)^{-1}  -1$
  as the time coordinate in the form 
      \begin{equation}
  ds^2 = H^{-2} (a) \left( \frac{da}{a}\right)^2 - a^2 dx^2 = \frac{1}{(1+z)^2}\left[ H^{-2} (z)dz^2 - dx^2\right]
  \end{equation} 
  This form clearly shows that the only dynamical content of the metric is encoded in the 
  function $H(a) = (\dot a/a)$. An immediate consequence is that any observation which is capable
  of determining the geometry of the universe can only provide --- at best --- information 
  about this function. 
  
 (b) Since cosmological observations usually use radiation received from distant sources,
  it is worth reviewing briefly the propagation of radiation in the universe. The radial
  light rays follow a trajectory given by 
    \begin{equation}
   r_{\rm em}(z) = S_k (\alpha); \qquad \alpha \equiv
   \frac{1}{a_0} \int_0^z H^{-1}(z) dz
  \end{equation} 
  if the photon is emitted at $r_{\rm em}$ at the redshift $z$ and received  here today.
  Two other quantities closely related to $r_{\rm em}(z)$ are the 
   luminosity distance, $d_L$, and the angular diameter distance $d_A$. If we receive a flux $F$ from a source of luminosity $L$, then the luminosity distance is defined via the relation \( F \equiv L/ 4\pi d_L^2(z) \). If an object of transverse length $l$ subtends a small angle $\theta$, the angular diameter distance is defined via (\( l = \theta d_A\) ). Simple calculation shows that:
     \begin{equation}
  d_L(z)  = a_0 r_{\rm  em}(z) (1+z) = a_0 (1+z) S_k(\alpha);
  \quad
  d_A(z)=a_0 r_{\rm em}(z)(1+z)^{-1}
  \label{dlzeqn}
  \end{equation} 

(c) As an example of determining the spacetime geometry of the universe  from observations, let us consider how one can determine $a(t)$ from the observations of the luminosity distance. It is clear from the 
first equation in \eq{dlzeqn} that 
    \begin{equation}
   H^{-1}(z) = \left[ 1 -  \frac{k d^2_L(z)}{a_0^2(1+z)^2}\right]^{-1/2} \frac{d}{dz}
  \left[  \frac{d_L(z)}{1+z}\right] \to \frac{d}{dz} \left[\frac{d_L(z)}{1+z}\right]
  \end{equation}
where the last form is valid for a $k=0$ universe. If we determine the form of $d_L(z)$ from observations --- which can be done if we can measure
the flux $F$ from a class of sources with known value for luminosity $L$ --- then we can use
this relation to determine the evolutionary history of the universe and thus the dynamics.

\section{Thermal History of the  Universe}

Let us next consider some key events in the evolutionary history of our universe \cite{tpsfuv3}.
The most well understood phase of the universe occurs when the temperature is less
than about $10^{12}$ K. Above this temperature, thermal production of baryons and 
their strong interaction is significant and somewhat difficult to model. We can ignore
such complications at lower temperatures and --- as we shall see --- several interesting
physical phenomena did take place during the later epochs with $T\lesssim 10^{12}$. 

The first thing we need to do is to determine the composition of the universe when
$T\approx 10^{12}$ K. We will certainly have, at this time, copious amount of photons and
all species of neutrinos and antineutrinos. In addition, neutrons and protons must exist
at this time since there is no way they could be produced later on. (This implies that
phenomena which took place at higher temperatures should have left a small excess of
baryons over anti baryons; we do not quite understand how this happened and will just
take it as an initial condition.) 
Since the rest mass of electrons correspond to a much lower temperature (about $0.5 \times 10^{10}$ K),
there will be large number of electrons and positrons at this temperature but in order to maintain
charge neutrality, we need to have a slight excess of electrons over positrons (by about 1 part
in $10^9$) with the net negative charge compensating the positive charge contributed by 
protons. 

 An elementary calculation using the known interaction rates show that all these particles are in thermal equilibrium at this epoch. Hence standard rules of 
statistical mechanics allows us to determine the number density ($n$), energy density ($\rho$) and
the pressure ($p$) in terms of the distribution function $f$:
 \begin{equation} 
 n=\int f({\bf k})d^3{\bf k}=
{g\over 2\pi^2}\int^{\infty}_m
{(E^2-m^2)^{1/2} EdE\over
\exp[(E-\mu)/T]\pm 1}
\end{equation}  
  \begin{equation}
  \rho=\int Ef({\bf k})d^3{\bf k}=
{g\over 2\pi^2}\int^{\infty}_m
{(E^2-m^2)^{1/2}E^2 dE\over
\exp[(E-\mu)/T]\pm 1}
\end{equation}   
 \begin{equation} 
 p= {1\over 3}\int d^3{\bf k}f({\bf k})kv({\bf k})=
\int{1\over 3} {|k|^2\over E}
f({\bf k})d^3{\bf k}={g\over 6\pi^2}\int^{\infty}_m
{(E^2-m^2)^{3/2} dE\over
e^{[(E-\mu)/T]}\pm 1}
\end{equation} 
Next, we can argue that the chemical potentials for electrons, positrons and neutrinos can be taken to be
zero.  For example, conservation of chemical potential in the reaction $e^+e^- \to 2 \gamma$ implies that
the chemical potentials of electrons and positrons must differ in a sign. But since the number
densities of electrons and positrons, which are determined by the chemical potential, are very close
to each other, the chemical potentials of electrons and positrons must be (very closely) equal to each other. Hence
both must be (very close to) zero. Similar reasoning based on lepton number shows that neutrinos should also
have zero chemical potential. Given this, one can evaluate the integrals for all the relativistic
species and we obtain for the total energy density
 \begin{equation}
  \rho_{{\rm total}}=\sum_{i={\rm boson}} g_i
\left({\pi^2\over 30}\right)T^4_i+
\sum_{i={\rm fermion}}{7\over 8}g_i
\left({\pi^2\over 30}\right)T^4_i=g_{{\rm total}}
\left({\pi^2\over 30}\right)T^4
\end{equation}
where
  \begin{equation} 
  g_{{\rm total}}\equiv\sum_{{\rm boson}} g_B+
\sum_{{\rm fermion}}{7\over 8} g_F.
\end{equation}
The corresponding entropy density is given by
  \begin{equation} 
  s\cong {1\over T}(\rho+p)=
{2\pi^2\over 45} qT^3; \qquad q\equiv q_{{\rm total}}=\sum_{{\rm boson}}g_B
+{7\over 8}
\sum_{{\rm fermion}} g_F.
\end{equation}

\subsection{Neutrino background}

As a simple application of the above result, let us consider the fate of neutrinos
in the expanding universe. From the standard weak interaction theory, one can compute
the reaction rate $\Gamma$ of the neutrinos with the rest of the species. When this
reaction rate fall below the expansion rate $H$ of the universe, the reactions cannot
keep the neutrinos coupled to the rest of the matter. A simple calculation \cite{tpsfuv3} shows that
the relevant ratio is given by
  \begin{equation}
   {\Gamma\over H}\simeq
\left({T\over 1.4{\rm MeV}}\right)^3=
\left({T\over 1.6\times 10^{10}{\rm K}}\right)^3
\end{equation}
Thus, for $T\lesssim 1.6 \times 10^{10}$ K, the neutrinos decouple from matter. At slightly
lower temperature, the electrons and positrons annihilate increasing the number density 
of photons. Neutrinos do not get any share of this energy since they have already decoupled
from the rest of the matter. As a result, the photon temperature goes up with respect to
the neutrino temperature once the $e^+e^-$ annihilation is complete. This increase in the temperature
is easy to calculate. As far as the photons are concerned,  the increase in the temperature   is essentially due to the change in the degrees of freedom $g$ and is given by:
 \begin{equation} \frac {(a T_{\gamma})^3_{\rm after}}{
(aT_{\gamma})^3_{\rm before}}=
\frac{g_{\rm before}}{g_{\rm after}}=\frac{\frac{7}{8}(2+2)+2}{2}
=\frac{11}{4}.
\label{photongast}
\end{equation}
(In the numerator, one 2 is for electron; one 2 is for positron; the $7/8$ factor arises because these are fermions. The final 2 is for photons. In the denominator, there are only photons to take care of.) Therefore 
\begin{eqnarray} 
 (aT_{\gamma})_{{\rm after}}  &= &
\left({11\over 4}\right)^{1/3}(aT_{\gamma})_{{\rm before}}= 
\left({11\over 4}\right)^{1/3}
(aT_{\nu})_{{\rm before}} \nonumber\\ 
& = & \left({11\over 4}\right)^{1/3}(aT_{\nu})_{{\rm after}}   {\simeq}
1.4 (aT_{\nu})_{{\rm after}}.
\end{eqnarray}
The first equality is from Eq.~(\ref{photongast}); the second arises because the photons and neutrinos had the same temperature originally; the third equality is from the fact that for decoupled neutrinos $aT_\nu$ is a constant. This result leads to the prediction that, at present, the universe will contain a bath of neutrinos which has temperature that is (predictably) lower than that of CMBR. The future detection of such a cosmic neutrino background will allow us to probe the universe at its earliest epochs.

\subsection{Primordial Nucleosynthesis}\label{sec:primenucleo}

 When the temperature of the universe is higher than the binding 
energy of the nuclei ($\sim$ MeV), none of the heavy elements (helium and the metals)
could have existed in the universe.
The binding energies of the first four light nuclei, $^2H$, $^3H$, $^3He$
and $^4He$ are $2.22\, {\rm MeV}$, $6.92\,{\rm MeV}$, $7.72\,{\rm MeV}$ and $28.3\,{\rm MeV}$ 
respectively.
This would  suggest that these nuclei could be
formed when the temperature of the universe is in the range of
$(1-30){\rm MeV}$. The actual synthesis takes place only at a much lower
temperature, $T_{\rm nuc}=T_n\simeq 0.1{\rm MeV}$. The main reason for this delay
is the `high entropy' of our universe, i.e., the high value for the
photon-to-baryon ratio, $\eta^{-1}$.
 Numerically, 
 \begin{equation}
 \eta = \frac{n_B}{n_\gamma} = 5.5\times 10^{-10} \left(\frac{\Omega_B h^2}{0.02}\right); \quad
 \Omega_Bh^2 = 3.65 \times 10^{-3} \frab{T_0}{2.73\ {\rm K}}^3 \eta_{10}
 \label{defofeta}
 \end{equation}
To see this, let us assume, for a moment, that the nuclear (and other) reactions are fast enough to
maintain thermal equilibrium between various species of particles and nuclei.
In thermal equilibrium, the number density of a nuclear species $^{A}N_Z$
with atomic mass $A$ and charge $Z$ will be
\begin{equation}
n_A=g_A
\left({m_A T\over 2\pi}\right)^{3/2}\exp
\left[-\left({m_A-\mu_A\over T}\right)\right].
\label{qnseden}
\end{equation}
From this one can obtain the equation
for the temperature $T_A$ at which the mass
fraction of a particular species-A will be of order unity $(X_A\simeq 1)$. We find that 
\begin{equation}
T_A\simeq{B_A/(A-1)\over \ln(\eta^{-1})+1.5\ln
(m_B/T)}
\label{qta}
\end{equation}
where $B_A$ is the binding energy of the species.
This temperature will be fairly lower than $B_A$
because of the large value of
$\eta^{-1}$.
For $^2H$, $^3He$ and
$^4He$ the value of $T_A$ is $0.07 {\rm MeV}$, $0.11 {\rm MeV}$ 
and $0.28 {\rm MeV}$ respectively. Comparison
with the binding energy of these nuclei shows that these values are
lower  than the corresponding binding energies $B_A$ 
by a factor of about 10, at least.

Thus, even when the thermal equilibrium is maintained, significant
synthesis of nuclei can occur only at $T\lesssim 0.3 {\rm MeV}$ and not
at higher temperatures.
If such is the case, then we would expect significant production
 $(X_A\lesssim 1)$ of 
nuclear species-A at temperatures $T\lesssim T_A$. It turns out, however, that
the rate of nuclear reactions is {\it not} high enough to maintain
thermal equilibrium between various species. We have to determine the
temperatures up to which thermal equilibrium can be maintained and redo the
calculations to find non-equilibrium mass fractions.
The general procedure for studying non equilibrium abundances in an expanding 
universe is based on  \textit{rate equations}. Since we will
require this formalism again in 
Section \ref{sec:decoupling} (for the study of recombination), we will
 develop it in a somewhat general context. 

Consider a reaction in which two particles 1 and 2 interact to form two
other particles 3 and 4. For example, $n+\nu_e \rightleftharpoons p+e$ constitutes one
such reaction which converts neutrons into protons in the forward direction
and protons into neutrons in the reverse direction; another example we will come 
across in the next section is $p+e\rightleftharpoons {\rm H}\ +\gamma$ where the forward
reaction describes recombination of electron and proton  forming a neutral hydrogen
atom (with the emission of a photon), while the reverse reaction is the photoionisation
of a hydrogen atom. In general, we are interested in how the number density $n_1$
of particle species 1, say, changes due to a reaction of the form $ 1+ 2 \rightleftharpoons
3+4$. 

We first note that even if there is no reaction, the number density will change
as $n_1 \propto a^{-3}$ due to the expansion of the universe; so what we are really after
is the change in $n_1 a^3$. Further, the forward reaction will be proportional to 
the product of the number densities $n_1 n_2$ while the reverse reaction will be proportional
to $n_3n_4$. Hence we can write an equation for the rate of change of particle species
$n_1$ as 
\begin{equation}
\frac{1}{a^3}\frac{d(n_1a^3)}{dt} = \mu(An_3n_4  - n_1 n_2).
\end{equation} 
The left hand side is the relevant rate of change over and above that due to the expansion
of the universe;
on the right hand side, the two proportionality constants have been written as 
$\mu$ and $(A\mu)$, both of which, of course, will be functions of time.
(The quantity $\mu$ has the dimensions of cm$^3$s$^{-1}$, so that $n\mu$ has the dimensions of
s$^{-1}$; usually $\mu\simeq \sigma v$ where $\sigma$ is the cross-section for the relevant process and $v$ is the relative velocity.)
The left hand side has to vanish
when the system is in thermal equilibrium with $n_i = n_i^{\rm eq}$, where
the superscript `eq' denotes the equilibrium densities for the different 
species labeled by $i=1 - 4$. This condition allows us to rewrite
$A$ as $A=n_1^{\rm eq} n_2^{\rm eq}/(n_3^{\rm eq} n_4^{\rm eq})$. Hence the 
rate equation becomes
\begin{equation}
\frac{1}{a^3}\frac{d(n_1a^3)}{dt} = \mu n_1^{\rm eq} n_2^{\rm eq}
\left(\frac{n_3n_4}{n_3^{\rm eq} n_4^{\rm eq}} - \frac{n_1n_2}{n_1^{\rm eq} n_2^{\rm eq}}
\right).
\label{master}
\end{equation}

In the left hand side, one can write $(d/dt) = H a (d/da)$ which shows that
the relevant time scale governing the process is $H^{-1}$. Clearly,
when $H/n\mu \gg 1$ the right hand side becomes ineffective because of the 
$(\mu/H)$ factor and the number of particles of species 1 does not change.
We  see that when the expansion rate of the universe is large compared
to the reaction rate, the given reaction is ineffective 
in changing the number of particles.
This certainly does \textit{not} mean that the reactions have reached thermal equilibrium
and $n_i = n_i^{\rm eq}$; in fact, it means exactly the opposite: The reactions
are not fast enough to drive the number densities towards equilibrium densities
and the number densities ``freeze out'' at non equilibrium values. Of course, 
the right hand side will also vanish when $n_i = n_i^{\rm eq}$ which is the 
other extreme limit of thermal equilibrium. 

Having taken care of the general  formalism, let us now 
apply it to the process  of nucleosynthesis
which requires protons and neutrons combining together to form bound nuclei of heavier
elements like deuterium, helium \textit{etc.}. The abundance of these elements
are going to be determined by the relative abundance of neutrons and protons
in the universe. Therefore, we need to first worry about the maintenance of thermal
equilibrium between protons and the neutrons in the early universe.
As long as 
the inter-conversion between $n$ and $p$ 
through the weak interaction processes $(\nu+ n\leftrightarrow p+e)$,
$(\overline e+n\leftrightarrow p+\overline\nu)$ and the `decay'
$(n\leftrightarrow p+e+\overline\nu)$, is rapid (compared to the expansion rate
of the universe), thermal equilibrium will be maintained. Then 
the
equilibrium
$(n/p)$ ratio will be
\begin{equation}
\left({n_n\over n_p}\right)=
{X_n\over X_p}=\exp(-Q/T),
\label{qbyt}
\end{equation}
where $Q=m_n-m_p = 1.293$ MeV. At high $(T\gg Q)$ temperatures, there will be equal
number of neutrons and protons but as the temperature drops below about 1.3 MeV, the 
neutron fraction will start dropping exponentially provided thermal equilibrium
is still maintained. To check whether thermal equilibrium is indeed maintained, we need to compare the expansion rate
with the reaction rate. The expansion rate is given by $H =(8\pi G \rho/3)^{1/2}$
where $\rho = (\pi^2/30) g T^4$ with $g \approx 10.75$ representing the effective
relativistic degrees of freedom present at these temperatures.
At $T = Q$, this gives $H \approx 1.1 $ s$^{-1}$. The reaction rate 
needs to be computed from weak interaction theory. The neutron to proton
conversion rate, for example, is well approximated
by 
\begin{equation}
\lambda_{np} \approx  0.29 \ {\rm s}^{-1} \ \left(\frac{T}{Q}\right)^5 \left[\left(\frac{Q}{T}\right)^2 + 6 \left(\frac{Q}{T}\right) + 12\right].
\end{equation} 
At $T=Q$, this gives $\lambda \approx  5$ s$^{-1}$, slightly more rapid
than the expansion rate. But as $T$ drops below $Q$, this decreases rapidly
and the reaction ceases to be fast enough to maintain thermal equilibrium.
Hence we need to work out the neutron abundance by using 
\eq{master}. 

Using $n_1 = n_n, n_3 = n_p$ and $n_2, n_4 = n_l$ where the subscript
$l$ stands for the leptons, \eq{master} becomes
\begin{equation}
\frac{1}{a^3}\frac{d(n_na^3)}{dt} = \mu n_l^{\rm eq}
\left(\frac{n_pn_n^{\rm eq}}{n_p^{\rm eq} } - n_n
\right).
\label{masterone}
\end{equation} 
We now use \eq{qbyt}, write $(n_l^{\rm eq}\mu) = \lambda_{np}$ which is the rate
for neutron to proton conversion and introduce the fractional abundance $X_n = n_n/(n_n+n_p)$.
Simple manipulation then  leads to the equation 
\begin{equation}
\frac{dX_n}{dt} = \lambda_{np} \left( (1 - X_n) e^{-Q/T} - X_n\right).
\label{xn}
\end{equation} 
Converting from the variable $t$ to the variable $s = (Q/T)$ and using
$(d/dt) = - HT(d/dT)$, the  equations we need to solve  reduce to 
\begin{equation}
- Hs \frac{dX_n}{ds} = \lambda_{np} \left( (1 - X_n) e^{-s} - X_n\right);
\quad H = (1.1\ {\rm sec}^{-1})\ s^{-4};\quad 
 \lambda_{np} = \frac{ 0.29 \ {\rm s}^{-1}}{s^5}  \left[s^2 + 6 s + 12\right].
\end{equation} 
 It is now straightforward to integrate these equations numerically and determine
 how the neutron abundance changes with time.  The neutron fraction falls out of equilibrium
 when temperatures drop below 1 MeV and it freezes to about 0.15 at temperatures
 below 0.5 MeV.

\begin{figure}
  \includegraphics[scale=0.5,angle=-90]{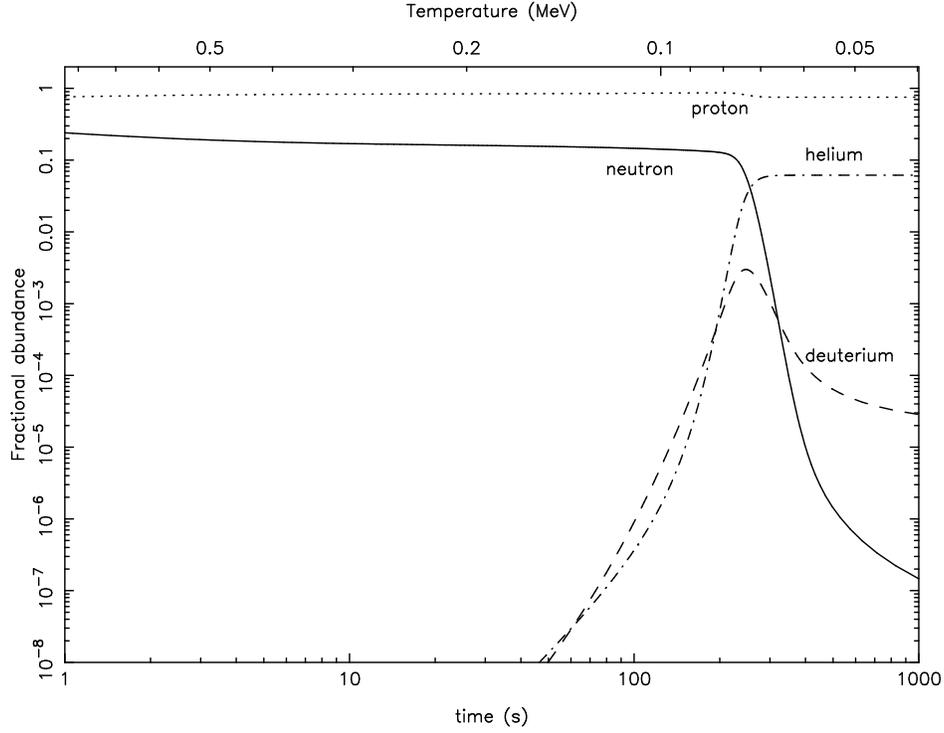} 
  \caption{The evolution of mass fraction of different species during nucleosynthesis}
  \label{fig:nucleo}
 \end{figure}

 As the temperature decreases further, the neutron decay with a half
 life of $\tau_n \approx 886.7$ sec (which is \textit{not} included 
 in the above analysis) becomes important and starts depleting
 the neutron number density.
The only way neutrons can survive is through the  
synthesis of light elements.
As the temperature falls further to $T=T_{\rm He}\simeq 0.28 {\rm MeV}$,
significant amount of He could have been produced if the nuclear
reaction rates were high enough. The possible reactions which produces ${}^4$He  are
  $[D(D,n)$ $^3He(D,p)$
$^4He$, $D(D,p)$ $^3H(D,n)$ $^4He, D(D,\gamma)$ $^4He]$.
These are all based on $D$, $^3He$ and $^3H$ and do not occur rapidly enough
because the mass fraction  of $D$, $^3He$ and $^3H$ are still quite 
small $[10^{-12}, 10^{-19}$ and $5\times 10^{-19}$ respectively]
at $T\simeq 0.3 {\rm MeV}$.
The reactions $n+p\rightleftharpoons
d+\gamma$ will lead to an equilibrium abundance ratio of deuterium
given by 
 \begin{equation}
 \fra{n_pn_n}{n_dn} = \fra{4}{3} \frab{m_pm_n}{m_d}^{3/2} \fra{\lb 2 \pi k_BT\rb^{3/2}}{\lb 2 \pi \hbar\rb^3n} e^{-B/k_BT}
  = \exp \left[ 25.82 - \ln \Omega_Bh^2 T_{10}^{3/2} - \frab{2.58}{T_{10}}\right].
 \label{qdeutabund}
 \end{equation}
 The equilibrium deuterium abundance passes through
unity (for $\Omega_B h^2 = 0.02$) at
 the temperature of about  $0.07$ MeV which is when the nucleosynthesis can really begin.

 So we  need to determine the neutron fraction at $T=0.07$ MeV given that
 it was about 0.15  at 0.5 MeV. During this epoch, the time-temperature relationship
 is given by $t= 130$ sec $(T/0.1$ MeV$)^{-2}$. The neutron  decay factor is $\exp(-t/\tau_n) \approx
 0.74$ for $T=0.07$ MeV. This decreases the neutron fraction to $0.15 \times 0.74 = 0.11$
 at the time of nucleosynthesis.  
When the temperature becomes $T\lesssim 0.07 {\rm MeV}$, the abundance of $D$ and
$^3H$ builds up and these elements further react to form $^4He$. A
good fraction of $D$ and $^3H$ is converted to $^4He$ (See Fig.\ref{fig:nucleo} which shows the growth of deuterium
and its subsequent fall when helium is built up). The resultant abundance 
of $^4He$ can be easily calculated by assuming that almost all neutrons end up
in $^4He$. Since each $^4He$ nucleus has two neutrons, $(n_n/2)$ helium nuclei 
can be formed (per unit volume) if the number density of neutrons is $n_n$.
Thus the mass fraction of $^4He$ will be
\begin{equation}
Y={4(n_n/2)\over n_n+n_p}=
{2(n/p)\over 1+(n/p)} = 2x_c
\end{equation}
 where $x_c =n/(n+p)$  is the neutron abundance at the time of production of deuterium.
  For $\Omega_Bh^2 = 0.02$, 
$x_c \approx 0.11$ giving $Y\approx 0.22$. Increasing baryon density to
$\Omega_Bh^2 =1$ will make $Y \approx 0.25$.
An accurate fitting formula for  the dependence of helium abundance on  
various parameters is given by
\begin{equation}
Y = 0.226 + 0.025 \log \eta_{10} + 0.0075 (g_* - 10.75) + 0.014 (\tau_{1/2} (n) - 10.3 \ {\rm min})
\end{equation}
where $\eta_{10}$ measures the baryon-photon ratio today via \eq{defofeta}
and $g_*$ is the effective number of relativistic degrees of freedom contributing to the energy density and $\tau_{1/2} (n)$ is the neutron half life.
The results (of a more exact treatment)
 are shown in \fig{fig:nucleo}.

As the reactions converting $D$ and $^3H$ to $^4He$ proceed, the number
density of $D$ and $^3H$ is depleted and the reaction rates - which
are proportional to $\Gamma\propto$ $X_A(\eta n_{\gamma})$
$<\sigma v>$ - become small. These reactions soon freeze-out leaving a
residual fraction of $D$ and $^3H$ (a fraction of about $10^{-5}$ to
$10^{-4}$). Since $\Gamma\propto\eta$ it is clear that the fraction of
$(D, ^3H)$ left unreacted will decrease with $\eta$. In contrast, the
$^4He$ synthesis - which is not limited by any reaction rate - is fairly
independent of $\eta$ and depends only on the $(n/p)$ ratio at
$T\simeq 0.1 {\rm MeV}$.
The best fits, with typical errors, to deuterium  abundance
 calculated from the theory, for the range 
$\eta = (10^{-10} - 10^{-9})$ is given by 
\begin{equation}
  Y_2 \equiv \frab{{\rm D}}{{\rm H}}_p = 3.6 \times 10^{-5\pm 0.06} \frab{\eta}{5\times 10^{-10}}^{-1.6}.
\end{equation}

The production of still heavier elements - even those like
$^{16}C$, $^{16}O$ which have \textit{higher} binding energies than $^4He$ - is 
suppressed in the early universe. Two factors are responsible for this
suppression:
(1) For nuclear reactions to proceed, the participating nuclei must
overcome their Coulomb repulsion. The probability to tunnel through
the Coulomb barrier is governed by the factor $F=\exp[-2 A^{1/3}$
$(Z_1 Z_2)^{2/3}(T/1 {\rm MeV})^{-1/3}]$
where $ A^{-1}=A^{-1}_1+A^{-1}_2$. For heavier nuclei (with larger $Z$), this factor
suppresses the reaction rate. 
(2) Reaction between helium and proton would have led to an element with atomic
   mass 5 while the reaction of two helium nuclei would have led to an element
   with atomic mass 8. However, there are no stable elements in the periodic
   table with the atomic mass of 5 or 8! 
   The ${}^8$Be, for example, has a half life of only $10^{-16}$ seconds.
   One can combine ${}^4$He with  ${}^8$Be to produce ${}^{12}$C but
   this can occur at significant rate only if it is a resonance reaction.
   That is, there should exist an excited state ${}^{12}$C nuclei which
   has an energy close to the interaction energy of ${}^4$He +  ${}^8$Be.
   Stars, incidentally, use this route to synthesize heavier elements. 
   It is this triple-alpha reaction
    which allows the synthesis of heavier elements
   in stars but it is not fast enough in the early universe.
   (You must thank your stars that  there is \textit{no} such resonance in ${}^{16}$O or in ${}^{20}$Ne
    ---  which is equally important for the survival of carbon and oxygen.)

The current observations indicate, with reasonable certainty 
that: (i) $(D/H)\gtrsim  1\times 10^{-5}$. (ii) $[(D+^3He)/H]$ $\simeq(1-8)\times 10^{-5}$  and (iii) $0.236< (^4He/H)$ $<0.254$. These observations are
consistent with the predictions if
 $10.3\,  \hbox{min}\lesssim \tau\lesssim 10.7\, \hbox{min}$, and
$
\eta=(3-10)\times 10^{-10}.
$
Using $\eta=2.68\times 10^{-8}\Omega_B h^2$, this leads to the important
conclusion:
$
0.011\leq \Omega_B h^2\leq 0.037.
$
When combined with the broad bounds on $h$,  $0.6\lesssim h\lesssim 0.8$, say, we can constrain the
baryonic density of the universe to be:
$
0.01\lesssim \Omega_B\lesssim 0.06.
$
These are the typical bounds on $\Omega_B$ available today. It shows
that, if $\Omega_{{\rm total}}\simeq 1$  then most of the matter in the universe must be non baryonic.

Since the $^4He$ production depends on $g$, the observed value of $^4He$
restricts
 the total energy density present at the time of nucleosynthesis. In particular, it constrains
 the number $(N_{\nu})$ of light neutrinos (that is, neutrinos with
$m_{\nu}\lesssim 1 {\rm MeV}$ which would have been relativistic at
$T\simeq 1 {\rm MeV}$). The observed abundance is best explained by $N_{\nu}=3$, is
barely consistent with $N_{\nu}=4$ and rules out $N_{\nu}>4$. The
laboratory bound on the total number of particles including
neutrinos, which  couples to
the $Z^0$ boson
 is determined by measuring the decay
width of the particle $Z^0$; each particle with mass less than $(m_z/2) \simeq 46 $ GeV contributes about
$180\ {\rm MeV}$ to this decay width. 
 This bound is $N_\nu = 2.79 \pm 0.63$ which is consistent with the
cosmological observations.

\subsection{Decoupling of matter and radiation}\label{sec:decoupling}
 
In the  early hot phase, the radiation will be in thermal equilibrium with
matter;  as the universe cools
below $k_BT \simeq (\epsilon_a/10)$ where $\epsilon_a$ is the 
binding energy of atoms, the electrons and ions will combine
 to form neutral atoms and radiation will decouple from matter.
 This occurs 
at $T_{\rm dec} \simeq 3\times 10^3$ K. 
As the universe expands further, these photons will continue to exist without any 
further interaction. It will retain thermal 
 spectrum since the redshift  of the frequency $\nu \propto a^{-1}$ is equivalent to changing the temperature in the spectrum by the scaling
 $T\propto (1/a)$.
It turns out that the major component of the extra-galactic
background light (EBL) which exists today is in the microwave band and can be fitted very accurately by
a thermal spectrum at a temperature of about $2.73\ {\rm K}$. 
It seems reasonable to interpret this radiation as a relic arising
from the early, hot, phase of the evolving universe.  
This relic radiation, called \textit{cosmic microwave background radiation}, turns out
to be a gold mine of cosmological information and is extensively investigated
in recent times. We shall now discuss some details related to  the formation of neutral atoms and the decoupling of photons.
\index{Cosmic Microwave Background Radiation}

The relevant reaction is, of course, $e+p \rightleftharpoons \mathrm{H} +\gamma$
and if the rate of this reaction is faster than the expansion rate, then one can calculate
the neutral fraction using Saha's equation.
Introducing the
fractional ionisation,
$X_i$,  for each of the particle species
and using the facts $n_p=n_e$ and $n_p+n_H=n_B$, it follows that
$X_p=X_e$ and $X_H=(n_H/n_B)$ $=1-X_e$. Saha's equation  now gives 
\begin{equation}
 {1-X_e\over X^2_e}
\cong 3.84 \eta(T/m_e)^{3/2}\exp (B/T)
\label{qrealsa}
\end{equation}
where $\eta=2.68\times 10^{-8}(\Omega_B h^2)$ is the baryon-to-photon ratio.
We may define $T_{\rm atom}$ as the temperature at which 90 percent of the
electrons, say,  have combined with protons: i.e. when $X_e=0.1$. This leads to
the condition:
\begin{equation}
(\Omega_B h^2)^{-1}\tau^{-{3\over 2}}\exp
\left[-13.6\tau^{-1}\right]=3.13\times 10^{-18}
\end{equation}
where $\tau=(T/1{\rm eV})$. For a given value of $(\Omega_B h^2)$, this
equation can be easily solved by iteration.
Taking logarithms and iterating once we find
$\tau^{-1}\cong 3.084-0.0735\ln (\Omega_B h^2)$
with the corresponding redshift $(1+z)=(T/T_0)$ given by
\begin{equation}
(1+z)=1367[1-0.024\ln(\Omega_B h^2)]^{-1}.
\end{equation}
For $\Omega_B h^2=1, 0.1,0.01$ we get $T_{\rm atom}$ 
$\cong 0.324{\rm eV}$, $0.307{\rm eV}$,
$0.292 {\rm eV}$ respectively. These values correspond to the redshifts of
$1367, 1296$ and $1232$.

Because the preceding analysis was based on equilibrium densities, it is
important to check that the rate of the reactions $p+e\leftrightarrow H+\gamma$
is fast enough to maintain equilibrium.
For $\Omega_B h^2 \approx 0.02$, the equilibrium condition is only marginally
 satisfied, making this analysis suspect. More importantly, the direct
 recombination to the ground state of the hydrogen atom --- which was used in
 deriving the Saha's equation --- is not very effective in producing neutral
 hydrogen in the early universe. The problem is that each such recombination 
 releases a photon of energy 13.6 eV which will end up ionizing another neutral
 hydrogen atom which has  been formed earlier. As a result, the direct recombination
 to the ground state does not change the neutral hydrogen fraction at the 
 lowest order. Recombination through the excited states of hydrogen is more 
 effective since such a recombination ends up emitting more than one photon
 each of which has an energy less than 13.6 eV.  Given these facts, it is necessary
 to once again use the rate equation developed in the previous section to track 
 the evolution of ionisation fraction. 
 
 A simple procedure for doing this, which captures the 
 essential physics, is as follows: 
 We again begin with \eq{master} and repeating the analysis done in the last section,
 now  with
 $n_1 = n_e, n_2 = n_p, n_3 = n_\mathrm{H}$ and $n_4 = n_\gamma$, 
 and defining $X_e = n_e/(n_e + n_\mathrm{H}) = n_p/ n_\mathrm{H}$
 one can easily derive the rate equation  for this case:
 \begin{equation}
\frac{dX_e}{dt} = \left[ \beta(1-X_e) - \alpha n_b X_e^2\right] 
= \alpha \left( \frac{\beta}{\alpha}(1-X_e) -  n_b X_e^2\right).
\label{xe}
\end{equation} 
This equation is analogous to \eq{xn}; the first term gives the photoionisation
rate which produces the free electrons and the second term is the recombination
rate which converts free electrons into hydrogen atom and we have used the 
fact $n_e = n_b X_e$ \textit{etc.}. Since we know that direct recombination to the ground
state is not effective, the recombination rate $\alpha$ is the rate for
capture of electron by a proton forming an excited state of hydrogen.
To a good approximation, this rate is given by 
\begin{equation}
\alpha = 9.78 r_0^2 c \left(\frac{B}{T}\right)^{1/2} \ln \left(\frac{B}{T}\right)
\end{equation} 
 where $r_0=e^2/m_ec^2$ is the classical electron radius. To integrate 
 \eq{xe} we also need to know $\beta/\alpha$. This is easy because
 in thermal equilibrium the right hand side of \eq{xe} should vanish
 and Saha's equation tells us the value of  $X_e$  in thermal equilibrium. 
 On using \eq{qrealsa}, this
 gives 
 \begin{equation}
\frac{\beta}{\alpha} = \left(\frac{m_e T}{2\pi}\right)^{3/2} \exp[-(B/T)].
\end{equation} 
We can now integrate \eq{xe} using the variable $B/T$ just as we used
the variable $Q/T$ in solving \eq{xn}. The result shows that the actual
recombination proceeds more slowly compared to that predicted by the 
Saha's equation.  
  The actual fractional ionisation is higher
than the value predicted by Saha's equation at temperatures below about
1300. For example, at $z=1300$, these values differ by a factor 3;
at $z\simeq 900$, they differ by a factor of 200. The value of 
$T_{\rm atom}$, however, does not change significantly.
A more rigorous analysis shows that, 
in the redshift range
of $800 < z<1200$, the fractional ionisation varies rapidly and is given (approximately)
by the formula,
\begin{equation}X_e=2.4\times 10^{-3}
{(\Omega_{\rm NR} h^2)^{1/2}\over (\Omega_B h^2)}
\left({z\over 1000}\right)^{12.75}.
\label{qfracion}
\end{equation}
This is obtained by fitting a curve to the numerical solution.

The formation of neutral atoms makes the photons decouple from the matter. The redshift for decoupling can be determined as the epoch at which the optical depth for photons is unity.
Using  \eq{qfracion}, we can compute the optical depth for
photons to be
\begin{equation}
 \tau=\int^t_0
n(t)X_e(t)\sigma_T dt=
\int^z_o n(z) X_e(z)\sigma_T
\left({dt\over dz}\right) dz
 \simeq 0.37 \left({z\over 1000}\right)^{14.25}
 \end{equation}
where we have used the relation $H_0 dt
\cong -\Omega_{\rm NR}^{-1/2} z^{-5/2} dz$
which is valid for $z\gg 1$.
This optical depth is unity at $z_{\rm dec}=1072$. 
From the optical depth, we can also compute the probability that the
photon was last scattered in the interval $(z, z+dz)$. This is given by
$(\exp-\tau)$ $(d\tau/dz)$ which can be expressed as
\begin{equation}
P(z)=e^{-\tau}{d\tau\over dz}=5.26\times 10^{-3}
\left({z\over 1000}\right)^{13.25}
\exp\left[-0.37\left({z\over 1000}\right)^{14.25}\right].
\label{photonscat}
\end{equation}
This $P(z)$ has a sharp maximum at $z\simeq 1067$ and a width of about
$\Delta z\cong 80$. It is therefore reasonable to assume that decoupling 
occurred at $z\simeq 1070$ in an interval of about $\Delta z\simeq 80$.
We shall see later  that the finite thickness of the
surface of last scattering has important observational consequences.

\section{Structure Formation and Linear Perturbation Theory}
  
Having discussed the evolution of the background universe, we now turn to the study of structure formation.
Before discussing the details, let us briefly summarise the broad picture
and give references to some of the topics that we will \textit{not} discuss.
The  key idea is that if there existed small fluctuations in the energy density in the early universe, then gravitational instability can amplify them in a well-understood manner  leading to structures like galaxies etc. today. The most popular model for generating these fluctuations is based on the idea that if the very early universe went through an inflationary phase \cite{inflation}, then the quantum fluctuations of the field driving the inflation can lead to energy density fluctuations\cite{genofpert,tplp}. It is possible to construct models of inflation such that these fluctuations are described by a Gaussian random field and are characterized by a power spectrum of the form $P(k)=A k^n$ with $n\simeq 1$ (see Sec. \ref{sec:inflation}). The models cannot predict the value of the amplitude $A$ in an unambiguous manner but it can be determined from CMBR observations. The CMBR observations are consistent with the inflationary model for the generation of perturbations and gives $A\simeq (28.3 h^{-1} Mpc)^4$ and $n=0.97\pm0.023$ (The first results were from COBE \cite{cobeanaly} and
WMAP has reconfirmed them with far greater accuracy).
When the perturbation is small, one can use well defined linear perturbation theory to study its growth. But when $\delta\approx(\delta\rho/\rho)$ is comparable to unity the perturbation theory
breaks down. Since there is more power at small scales, smaller scales go non-linear first and structure forms hierarchically. 
The non linear evolution of the  \textit{dark matter halos} (which is an example of statistical mechanics
 of self gravitating systems; see e.g.\cite{smofgs}) can be understood by simulations as well as theoretical models based on approximate ansatz
\cite{nlapprox} and  nonlinear scaling relations \cite{nsr}.
 The baryons in the halo will cool and undergo collapse
 in a fairly complex manner because of gas dynamical processes. 
 It seems unlikely that the baryonic collapse and galaxy formation can be understood
 by analytic approximations; one needs to do high resolution computer simulations
 to make any progress \cite{baryonsimulations}. 
 All these results are broadly consistent with observations. 

 As long as these fluctuations are small, one can study their evolution by linear perturbation theory, which is what we will start with
 \cite{linpertpeda}.
 The basic idea of linear perturbation theory is well defined and simple. We perturb the background FRW metric by
$g_{ik}^{FRW}\to g_{ik}^{FRW}+h_{ik}$ and also perturb the source energy momentum tensor by
$T_{ik}^{FRW}\to T_{ik}^{FRW}+\delta T_{ik}$. Linearising the Einstein's equations, one can relate the perturbed quantities by a relation of the form
${\cal L}( g_{ik}^{FRW} )h_{ik}=\delta T_{ik}$ 
where  ${\cal L}$ is second order linear differential operator depending on the back ground metric  $g_{ik}^{FRW}$ .
Since the background is maximally symmetric, one can separate out time and space;
for e.g, if $k=0$, simple Fourier modes can be used for this purpose and we can write down the equation for any given mode, labelled by a wave vector $\mathbf{k}$ as:
\begin{equation} 
{\cal L}(a(t),{\bf k})h_{ab}(t,{\bf k})=\delta T_{ab}(t,{\bf k})
\end{equation} 
To every mode we can associate a wavelength normalized to today's value:
$\lambda(t)=(2\pi/k)(1+z)^{-1}$ and a corresponding mass scale 
 which is invariant under expansion:
\begin{equation} 
M={4\pi \rho(t)\over 3} \left[{\lambda(t)\over 2}\right]^3 
{=}
    {4\pi \rho_0\over 3}  \lb {\lambda_0\over 2}\rb^3    
{ = }
  1.5\times 10^{11}{\rm M}_\odot (\Omega_{m} h^2) \lb {\lambda_0\over 1\, {\rm Mpc}}\rb^3.
 \end{equation}
The behaviour of the mode depends on the relative value of $\lambda(t)$ as compared to the Hubble radius
$d_H(t)\equiv (\dot a/a)^{-1}$. Since the Hubble radius: $d_H(t)\propto t$ while the 
wavelength of the mode: $\lambda(t)
\propto a(t)\propto (t^{1/2},t^{2/3})$ in the radiation dominated and matter dominated phases it follows that $\lambda(t)>d_H(t)$ at sufficiently early times. When $\lambda(t)=d_H(t)$, we say that the mode is entering the Hubble radius. Since the 
  Hubble radius at \( z= z_{\rm eq} \) is
 \begin{equation} 
 \lambda_{\rm eq} \cong   \frab{H_0^{-1}} { \sqrt 2} \frab{\Omega_R^{1/2} }{ \Omega_{NR}}  \cong 14 {\rm Mpc}   (\Omega_{\rm NR} h^2)^{-1}  
 \end{equation}
it follows that modes with
 $ \lambda_0 > \lambda_{\rm eq} $ enter Hubble radius in MD phase while
 the more  relevant modes with $ \lambda < \lambda_{\rm eq} $  enter in the RD phase.   
 Thus, for a given mode we can identify three distinct   phases: First, very early on,  when
$\lambda>d_H, z>z_{eq}$  the dynamics is described by general relativity. In this stage, the universe is radiation dominated,  gravity is the only relevant force and the perturbations are linear.   Next, when
$\lambda<d_H$ and $ z>z_{eq}$ one can describe the dynamics by   Newtonian considerations. The perturbations are still linear and the universe is  radiation dominated. Finally, when 
$\lambda<d_H, z<z_{eq}$ we have a matter dominated universe in which we can use the  Newtonian formalism;
but at this stage --- when most astrophysical structures form --- we need to grapple with  nonlinear
astrophysical processes.

Let us now consider the metric perturbation  in greater detail. When the metric is perturbed to the form: 
$g_{ab} \to g_{ab} + h_{ab}$ the  perturbation can be split as $h_{ab} = (h_{00},h_{0\alpha}\equiv w_\alpha,h_{\alpha\beta})$. We also know that any 
 3-vector ${\bf w}(\textbf{x})$ can be split as $\textbf{w} = \textbf{w}^\perp + \textbf{w}^\parallel$
 in which  $\textbf{w}^\parallel = \nabla \Phi^\parallel$ is curl-free (and  carries one degree of freedom)   while $\textbf{w}^\perp$ is divergence-free (and has 2 degrees of freedom). This result is obvious in  $\textbf{k}-$space since we can write any vector
 $ \textbf{w}(\textbf{k})$ as a sum of two terms, one along $\mathbf{k}$ and one transverse to $\mathbf{k}$:
   \begin{equation}
   \textbf{w}(\textbf{k}) = \textbf{w}^\parallel (\textbf{k}) + \textbf{w}^\perp (\textbf{k})= \underbrace{\textbf{k} \left( \frac{\textbf{w}(\textbf{k}) \cdot \textbf{k}}{k^2}\right)}_{\rm along\ {\bf k}} + \underbrace{\left[ \textbf{w}(\textbf{k}) - \textbf{k} \left( \frac{\textbf{k} \cdot \textbf{w}(\textbf{k})}{k^2}\right)\right]}_{{\rm transverse\ to \ }{\bf k}} ; \ \textbf{k} \times \textbf{w}^\parallel = 0; \ \textbf{k}\cdot \textbf{w}^\perp =0
   \end{equation}
   Fourier transforming back, we can split $\mathbf{w}$ into a curl-free and divergence-free parts.
 Similar decomposition works for $h_{\alpha\beta}$ by essentially repeating the above analysis on each index. We can write: 
 \begin{equation}
 h_{\alpha\beta} = \underbrace{\psi \delta_{\alpha\beta}}_{\rm trace} + \underbrace{\left( \nabla_\alpha u_\beta^\perp +  \nabla_\beta u_\alpha^\perp\right)}_{\rm traceless\ from\ vector} + \underbrace{\left(\nabla_\alpha\nabla_\beta - \frac{1}{3} \delta_{\alpha\beta}\nabla^2\right)\Phi_1}_{\rm traceless\ from\ scalar}
 + h_{\alpha\beta}^{\perp\perp} \Rightarrow 1 + 2 + 1 + 2 = 6
 \end{equation} 
 The $u_\alpha^\perp$ is divergence free and $h_{\alpha\beta}^{\perp\perp}$ is traceless and divergence free.
 Thus the most general perturbation $h_{ab}$ (ten degrees of freedom) can be built out of
 \begin{equation}
h_{ab} = (h_{00},h_{0\alpha}\equiv w_\alpha,h_{\alpha\beta})=[h_{00},(\Phi^\parallel,\textbf{w}^\perp), 
(\psi, \Phi_1,u_\alpha^\perp,h_{\alpha\beta}^{\perp\perp})] \Rightarrow[1,(1,2),(1,1,2,2)]
\end{equation} 
 We now use the freedom available in the choice of
 four  coordinate transformations  to set four conditions: $\Phi^\parallel = \Phi_1 =0$ and $u_\alpha^\perp =0$ thereby
 leaving six degrees of freedom in $(h_{00} \equiv  2\Phi, \psi, \textbf{w}^\perp, h_{\alpha\beta}^{\perp\perp})$ as nonzero.
Then the perturbed line element takes the form:
\begin{equation}
\de s^2 = a^2(\eta) ~ [\{1 + 2 \Phi(\xeta) \} ~ \de \eta^2 
-2 w_{\alp}^\perp(\xeta) ~ \de \eta ~ \de x^{\alp}
- \{(1 - 2 \psi(\xeta)) ~ \delta_{\alp \beta} 
+ 2 h_{\alp\beta}^{\perp\perp}(\xeta) \} ~ \de x^{\alp} ~ \de x^{\beta}]
\end{equation} 
To make further simplification we need to use two facts from Einstein's equations. It turns out that the
Einstein's equations for $\textbf{w}^\perp$ and $ h_{\alpha\beta}^{\perp\perp}$ decouple from those
 for ($\Phi, \psi$).  Further, in the absence of anisotropic stress, one of the equations give $\psi = \Phi$. If we use these two facts, we can simplify the structure of perturbed metric drastically. As far as the growth of matter perturbations are concerned, we can ignore $w_\alpha^\perp$ and $h_{\alpha\beta}^{\perp\perp}$  and work with a simple metric:
 \begin{equation} 
 ds^2 = a^2(\eta)[(1+2\Phi)d\eta^2 - (1-2\Phi)\delta_{\alpha\beta}dx^{\alpha}dx^{\beta}]
 \label{pertmetricfinal}
 \end{equation}
with just one perturbed scalar degree of freedom in $\Phi$. This is what we will study.

Having decided on the gauge, let us consider the evolution equations for the perturbations. While one can directly work with the Einstein's equations, it turns out to be convenient to use the equations of motion for matter variables, since we are eventually interested in the matter perturbations. In what follows, we will use the over-dot to denote $(d/d\eta)$ so that the standard Hubble parameter is $H=(1/a)(da/dt) = \dot a/a^2$. With this notation, the
continuity  equation becomes:
\begin{equation}
\dot \rho + 3 \left( a H - \dot \Phi\right) (\rho + p) = - \nabla_\alpha \left[ (\rho + p) v^\alpha \right]
\end{equation}
Since the momentum flux in the relativistic case is $(\rho+p)v^\alpha$, all the terms in the above equation are intuitively obvious, except probably the $\dot \Phi$ term. To see the physical origin of this term, note that the perturbation in \eq{pertmetricfinal} changes  the factor in front of the spatial metric from
 $a^2$ to $a^2(1 - 2\Phi)$ so that $ \ln a\to \ln a - \Phi$; hence the effective Hubble parameter
  from $(\dot a/a)$ to $(\dot a/a) -\dot \Phi$ which explains the extra $\dot \Phi$ term. 
 This is, of course, the exact equation for matter variables in the perturbed metric given by \eq{pertmetricfinal}; but we only need terms which are of linear order. Writing the curl-free velocity part as  $v^\alpha=\nabla^\alpha v$, the \textit{linearised} equations, for dark matter (with $p=0$) and radiation (with $p =  (1/3)\rho$)  perturbations are given by:
\begin{equation}
\dot \delta_m = \frac{d}{d\eta} \left( \frac{\delta n_m}{n_m}\right) = \nabla^2 v_m + 3 \dot \Phi;\quad  
\frac{3}{4} \dot\delta_R = \frac{d}{d\eta} \left( \frac{\delta n_R}{n_R}\right)=\nabla^2 v_R + 3\dot \Phi
\label{fifty}
\end{equation}
where $n_m$ and $n_R$ are the number densities of dark matter particles and radiation.
The same equations in Fourier space [using the same symbols for, say, $\delta(t,{\bf x})$ or $\delta(t,{\bf k})$] are simpler to handle:
\begin{equation}
\dot \delta_m = \frac{d}{d\eta} \left( \frac{\delta n_m}{n_m}\right)=-k^2 v_m + 3 \dot \Phi; 
\quad 
\frac{3}{4} \dot\delta_R = \frac{d}{d\eta} \left( \frac{\delta n_R}{n_R}\right)= -k^2 v_R + 3 \dot \Phi
\label{fiftyone}
\end{equation}
Note that these equations imply
\begin{equation}
\frac{d}{d\eta}\left[ \frac{\delta n_R}{n_R} - \frac{\delta n_m}{n_m}\right] = \frac{d}{d\eta} \left[ \delta \ln \left( \frac{n_R}{n_m}\right) \right] = \frac{d}{d\eta} \left[ \delta \left( \ln \left(\frac{s}{n_m}\right)\right) \right] = - k^2 ( v_R- v_m)
\label{vrminusvm}
\end{equation}
For long wavelength perturbations (in the limit of $k\to0$), this will lead to the conservation of perturbation $\delta(s/n_m)$ in  the entropy per particle.

Let us next consider the  
Euler equation which has the general form:
\begin{equation}
\partial_\eta [(\rho +p) v^\alpha] = - (\rho+p)\nabla^\alpha \Phi - \nabla^\alpha p - 4 a H (\rho+p) v^\alpha
\end{equation}
Once again each of the terms is simple to interpret. The $(\rho+p)$ arises because the pressure also contributes to inertia in a relativistic theory and the factor 4 in the last term on the right hand side arises because the term $v^\alpha \partial_\eta (\rho+p)$ on the left hand side needs to be compensated. Taking the linearised limit of this equation, for
dark matter and radiation, we get: 
\begin{equation}
\dot v_m = \Phi - aH v_m; \quad \dot v_R = \Phi + \frac{1}{4} \delta_R
\label{fiftythree}
\end{equation}

Thus we now have four equations in \eqs{fiftyone}, (\ref{fiftythree}) for the five variables $(\delta_m,\delta_R,v_m,v_R,\Phi)$. 
All we need to do is to pick
one more from
Einstein's equations to complete the set. The Einstein's equations for our perturbed metric are:
\begin{eqnarray}
^0_0 \ \mathrm{component} &:& k^2  \Phi 
+ 3 \frac{\dot{a}}{a} \left(\dot{\Phi} +  \frac{\dot{a}}{a}\Phi\right)  =  - 4 \pi G a^2 \sum_A \rho_A \delta_A=
- 4 \pi G a^2 \rho_{bg}\delta_{total}\label{fiftyfive}
\\ 
^0_{\alp} \ \mathrm{component} &:& \dot{\Phi} + \frac{\dot{a}}{a} \Phi =
- 4 \pi G a^2 \sum_A( \rho + p )_A v_A;\quad  \mathbf{v}=\nabla v\label{fiftysix}
\\ 
^{\alp}_{\alp} \ \mathrm{component} &:& 3 \frac{\dot{a}}{a} \dot{\Phi} + 2 \frac{\ddot{a}}{a} \Phi - 
\frac{\dot{a}^2}{a^2} \Phi + \ddot{\Phi} 
= 4 \pi G a^2    \delta p
\end{eqnarray}
where $A$ denotes different components like dark matter, radiation etc. Using \eq{fiftysix} in \eq{fiftyfive} we can get a modified  Poisson equation which is purely algebraic: 
\begin{equation} 
-k^2 \Phi = 4 \pi G a^2 \sum_A \left( \rho_A \delta_A - 3 \left( \frac{\dot a}{a}\right) (\rho_A + p_A) v_A\right)
\end{equation}
which once again emphasizes the fact that in the relativistic theory, both pressure and density act as source of gravity.

To get a feel for the solutions let us consider a flat universe dominated by a
  single component of matter with  the equation of state  $  p  = w  \rho  $. (A purely radiation dominated universe, for example, will have $w=1/3$.)
In this case the Friedmann background equation gives 
$\rho \propto a^{-3 (1+w)}$
and
\begin{equation}
\frac{\dot{a}}{a} = \frac{2}{(1+3 w) \eta}; ~~~
\frac{\ddot{a}}{a} = \frac{2 (1-3 w)}{(1+3 w)^2 \eta^2} 
\end{equation}
The equation for the potential $\Phi$ can be reduced to the form: 
\begin{equation}
\ddot{\Phi} + \frac{6 (1 + w)}{1 + 3 w} \frac{\dot{\Phi}}{\eta} 
+ k^2 w \Phi = 0
\label{poteqn}
\end{equation}
The second term is the damping due to the expansion while last term is the pressure support that will lead to oscillations. Clearly, the factor $k\eta$ determines which of these two terms dominates. When the pressure term dominates $(k\eta\gg 1$), we expect oscillatory behaviour while
when the background expansion dominates $(k\eta\ll 1$), we expect the growth to be suppressed. This is precisely what happens.
The exact solution is given in terms of the Bessel functions
\begin{equation}
\Phi(\eta)=\frac{C_1({\bf k}) J_{\nu/2}(\sqrt{w} k \eta) 
+ C_2({\bf k}) Y_{\nu/2}(\sqrt{w} k \eta)}{\eta^{\nu/2}}; \quad \nu = \frac{5 + 3 w}{1 + 3 w}
\label{sixtytwo}
\end{equation}
From the theory of Bessel functions, we know that:
\begin{equation}
\lim_{x \to 0} J_{\nu/2}(x) \simeq \frac{x^{\nu/2}}{2^{\nu/2} \Gamma(\nu/2 +1)}; \quad
\lim_{x \to 0} Y_{\nu/2}(x) \propto - \frac{1}{x^{\nu/2}}
\end{equation} 
This shows that if we want a finite value for $\Phi$ as $\eta \to 0$, we can set $C_2 =0$.
This gives the gravitational potential to be
\begin{equation}
\Phi(\eta)=\frac{C_1({\bf k}) J_{\nu/2}(\sqrt{w} k \eta)}{\eta^{\nu/2}}; \quad \nu = \frac{5 + 3 w}{1 + 3 w}
\label{sixtya}
\end{equation}
 The corresponding density perturbation will be: 
\begin{equation}
\delta  = -2\Phi-\frac{(1 + 3 w)^2 k^2 \eta^2}{6} 
\frac{C_1({\bf k}) J_{\nu/2}(\sqrt{w} k \eta) }{\eta^{\nu/2}} 
+ (1 + 3 w) 
\sqrt{w} k \eta \frac{C_1({\bf k}) J_{(\nu/2)+1}(\sqrt{w} k \eta) }{\eta^{\nu/2}}
\label{sixthree}
\end{equation}
To understand the nature of the solution, note that 
$ d_H = (\dot a / a)^{-1} \propto \eta$ and $ \quad kd_H \simeq d_H/\lambda \propto k\eta$. So the argument of the Bessel function is just the ratio $(d_H/\lambda)$. 
From the theory of Bessel functions, we know that for small values of the argument $J_\nu(x) \propto x^\nu$ is a power law while for large values of the argument it oscillates with a decaying amplitude:
\begin{equation}
\lim_{x \to \infty} J_{\nu/2}(x) \sim \frac{\cos[x-(\nu-1) \pi/4]}{\sqrt{x}}; 
\end{equation}
Hence, for modes which are still outside the Hubble radius  ($k \ll \eta^{-1}$), we have 
a constant amplitude for the potential and density contrast:
\begin{equation}
\Phi \approx  \Phi_i (\mathbf{k}); \quad \delta  \approx -2\Phi_i(\mathbf{k})
\end{equation}
That is, the
perturbation is frozen (except for a decaying mode) at a constant value. On the other hand,
for modes which are inside the Hubble radius ($k \gg \eta^{-1}$), the perturbation is  rapidly oscillatory (if $w\ne 0$). That is the pressure is effective at small scales and leads to acoustic oscillations in the medium. 

A special case of the above is the flat,
matter-dominated universe with $w=0$. In this case, we need to take the $w\to 0$ limit
 and 
 the general solution is indeed 
 a constant $\Phi = \Phi_i(\mathbf{k})$ (plus  a decaying mode   $\Phi_{decay} \propto \eta^{-5} $ which diverges as $\eta \to 0$).
The corresponding density perturbations is: 
\begin{equation}
\delta = -(2 + \frac{k^2 \eta^2}{6})\Phi_i(\mathbf{k})
\label{six9}
\end{equation} 
which shows that density perturbation is ``frozen" at large scales  but grows at small scales:
\begin{equation}
\delta = 
\begin{cases}
-2\Phi_i(\mathbf{k}) = {\rm constant}\quad \text{$(k\eta \ll 1)$}\\
- \frac{1}{6} k^2 \eta^2 \Phi_i(\mathbf{k}) \propto \eta^2 \propto a \quad\text{$(k\eta \gg 1)$}
\end{cases}
\label{sixtynine}
\end{equation}
We will use these results later on.

 \section{ Perturbations in Dark Matter and Radiation}
  
We shall now move on to the more realistic case of a multi-component universe consisting of radiation and collisionless dark matter.
(For the moment we are ignoring the baryons, which we will study in Sec. \ref{sec:tempcmbr}). It is convenient to use $y=a/a_{eq}$ as independent variable
rather than the time coordinate. The background expansion of the universe described by the function $a(t)$ can be equivalently expressed (in terms of the conformal time $\eta$) as 
\begin{equation}  
y \equiv \frac{\rho_M }{\rho_R} = \frac{a}{a_{\rm eq}} = x^2 + 2x, \qquad x\equiv \left(\frac{\Omega_M }{4a_{\rm eq}}\right)^{1/2} H_0\eta
\end{equation} 
It is also useful to define a critical  wave number $k_c$ by:
\begin{equation}
k_c^2 = \frac{ H_0^2 \Omega_m }{ a_{\rm eq}} = 4(\sqrt{2} - 1)^2 \eta_{\rm eq}^{-2}= 4(\sqrt{2} - 1)^2 k_{\rm eq}^2; \qquad k_c^{-1}=19(\Omega_m h^2)^{-1}Mpc
\end{equation}
which essentially sets the comoving scale corresponding to matter-radiation equality. Note that $2x=k_c\eta$ and $y\approx k_c\eta$ in the radiation dominated phase while $y=(1/4)(k_c\eta)^2$ in the matter dominated phase.

We now manipulate \eqs{fiftyone}, (\ref{fiftythree}), (\ref{fiftyfive}), (\ref{fiftysix}) governing the growth of perturbations by essentially eliminating the velocity. This leads to the three equations 
  \begin{equation}
   y \Phi' + \Phi + \frac{1}{3} \frac{k^2}{k_c^2} \frac{y^2}{1+y} \Phi = - \frac{1}{2} \frac{y}{1+y} \left( \delta_m + \frac{1}{y} \delta_R\right)
   \label{first}
   \end{equation}
   \begin{equation}
   (1+y) \delta_m'' +  \frac{2+3y}{2y} \delta_m' = 3 (1+y) \Phi'' + \frac{3(2+3y)}{2y} \Phi' - 
   \frac{k^2}{k_c^2}  \Phi
   \label{sec}
   \end{equation}
   \begin{equation}
   (1+y) \delta_R'' + \frac{1}{2}  \delta_R' +\frac{1}{3} \frac{k^2}{k_c^2}\delta_R = 4 (1+y) \Phi'' + 2 \Phi' - \frac{4}{3} \frac{k^2}{k_c^2} \Phi
   \label{three}
   \end{equation}
for the three unknowns $\Phi,\delta_m,\delta_R$. Given suitable initial conditions we can solve these equations to determine the growth of perturbations. The initial conditions need to imposed very early on when the modes are much bigger than the Hubble radius which corresponds to the   $y\ll 1, k\to 0$ limit. In this limit, the equations become:
\begin{equation}
   y \Phi' + \Phi\approx - \frac{1}{2}  \delta_R;\quad
   \delta_m'' +  \frac{1}{y} \delta_m' \approx 3  \Phi'' + \frac{3}{y} \Phi';\quad 
  \delta_R'' + \frac{1}{2}  \delta_R' \approx 4  \Phi'' + 2 \Phi' 
  \label{yto0}
   \end{equation}
   We will take $\Phi(y_i,k)=\Phi_i(k)$ as given value, to be determined by the processes that generate the initial perturbations. First equation in \eq{yto0} shows that we can take $\delta_R=-2\Phi_i$ for $y_i\to0$. Further \eq{vrminusvm} shows that adiabaticity is respected at these scales and we can take 
 $
 \delta_m=(3/4)\delta_R=-(3/2)\Phi_i;
$. The exact equation \eq{first} determines $\Phi'$ if $(\Phi,\delta_m,\delta_R)$ are given. Finally we use
the last two  equations to set $\delta\,'_m=3\Phi\,',\delta\,'_R=4\Phi\,'$, Thus we take the initial conditions at some $y=y_i\ll1$ to be:
\begin{equation}
\Phi(y_i,k)=\Phi_i(k);\quad
\delta_R(y_i,k)=-2\Phi_i(k);\quad
\delta_m(y_i,k)=-(3/2)\Phi_i(k)
\end{equation} 
with 
$\delta\,'_m(y_i,k)=3\Phi\,'(y_i,k);\
\delta\,'_R(y_i,k)=4\Phi\,'(y_i,k)$.

\begin{figure}[ht]
\begin{center}
\includegraphics[scale=0.65]{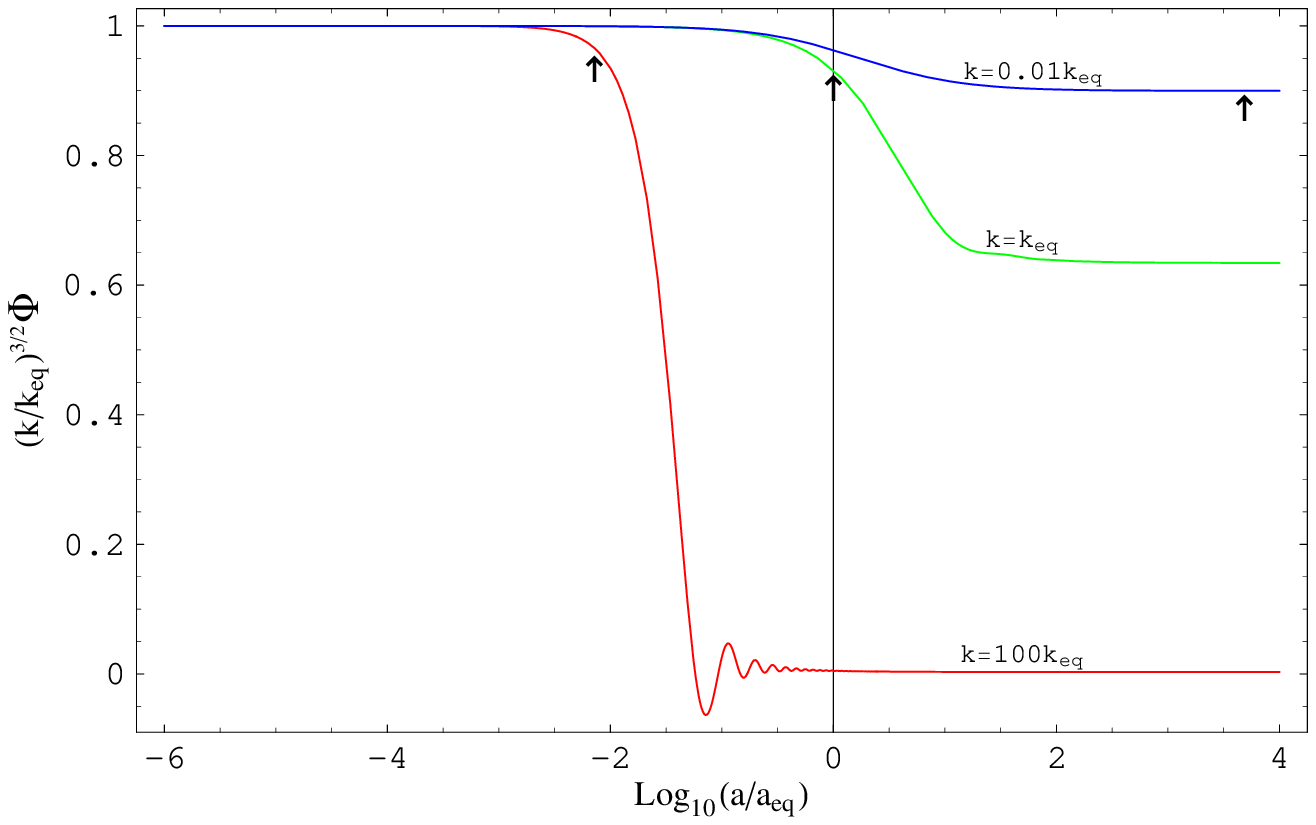}\quad\includegraphics[scale=0.75]{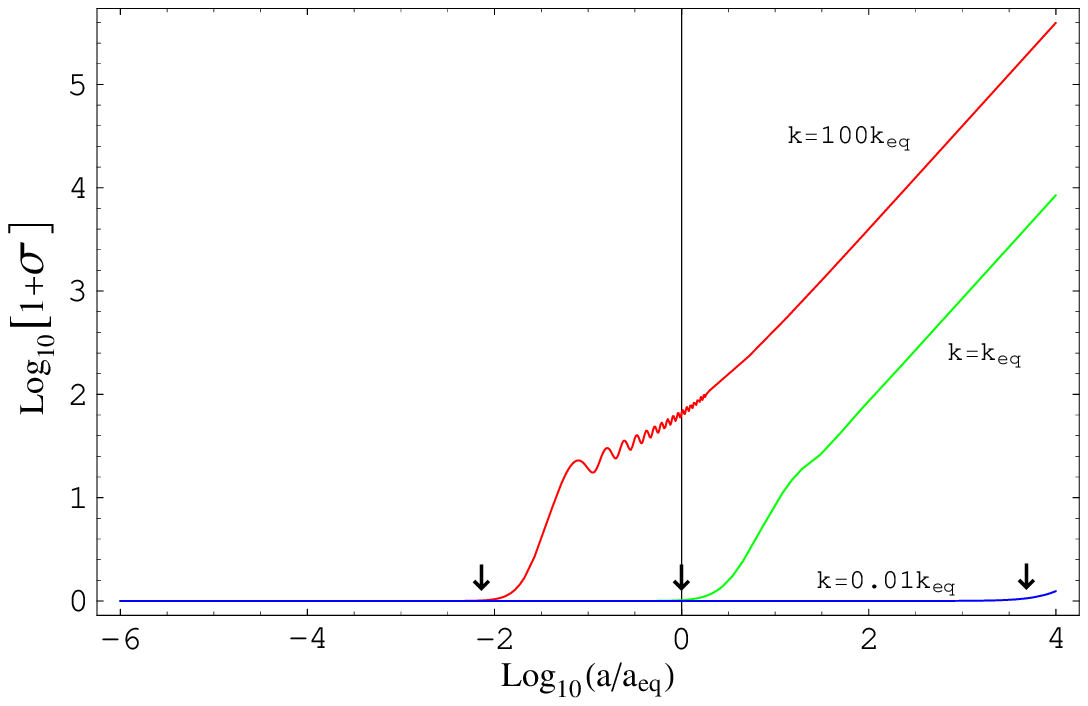}
\end{center}
\caption{Left Panel:The evolution of gravitational potential $\Phi$ for 3 different modes. The wavenumber
is indicated by the label and the epoch at which the mode enters the Hubble radius
is indicated by a small arrow. The top most curve is for a mode which stays outside the 
Hubble radius for most  of its evolution and is well described by \eq{exactsol}. The other two 
modes show the decay of $\Phi$ after the mode has entered the Hubble radius in the radiation
dominated epoch as described by \eq{Phieqn}. Right Panel: Evolution of entropy perturbation (see \eq{defofent} for the definition). The entropy perturbation is essentially zero till the mode enters Hubble radius and grows afterwards tracking the dominant energy density perturbation. }
\label{fig:phi}
\end{figure}

\begin{figure}[hb]
\begin{center}
\includegraphics[scale=0.8]{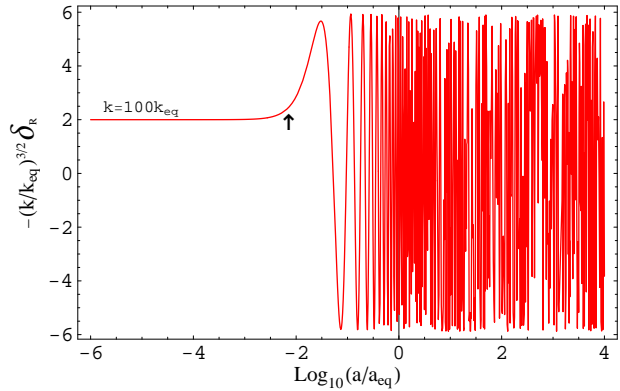}
\end{center}
\caption{Evolution of $\delta_R$ for a mode with  $k= 100 k_{\rm eq}$.
The mode remains frozen outside the Hubble radius at $ (k/k_{eq})^{3/2}(-\delta_R) \approx (k/k_{eq})^{3/2} 2\Phi=2$ (in the normalisation used
in Fig.~\ref{fig:phi} ) and 
oscillates when it enters the Hubble radius. The oscillations are well 
described by \eq{rados} with an amplitude of $6$. }
\label{fig:dr1}
\end{figure}

\begin{figure}[ht]
\begin{center}
\includegraphics[scale=0.8]{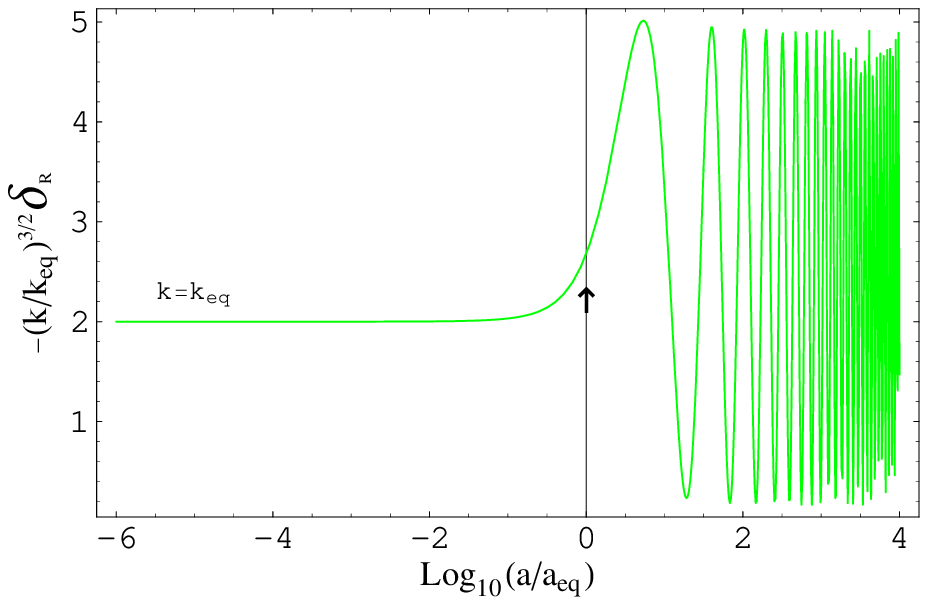}\quad\includegraphics[scale=0.8]{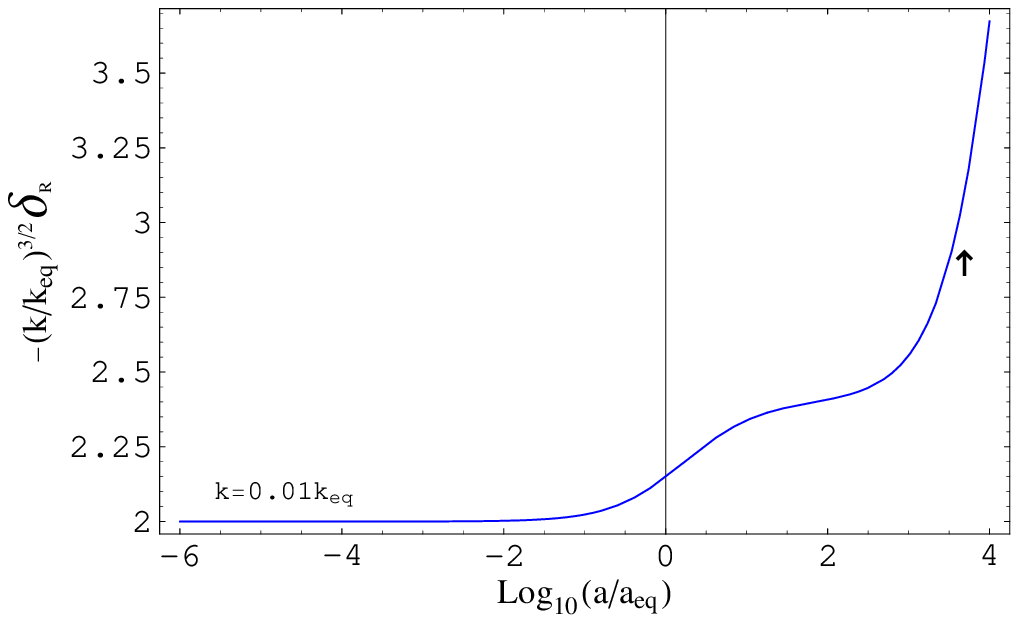}
\end{center}
\caption{Evolution of $\delta_R$ for two modes $k= k_{\rm eq}$ and $k= 0.01\, k_{\rm eq}$.
The modes remain frozen outside the Hubble radius at $(-\delta_R) \approx 2$ and 
oscillates when it enters the Hubble radius. The mode in the right panel stays outside the 
Hubble radius for most part of its evolution and hence changes very little.}
\label{fig:dr}
\end{figure}

\begin{figure}[hb]
\begin{center}
\includegraphics[scale=0.8]{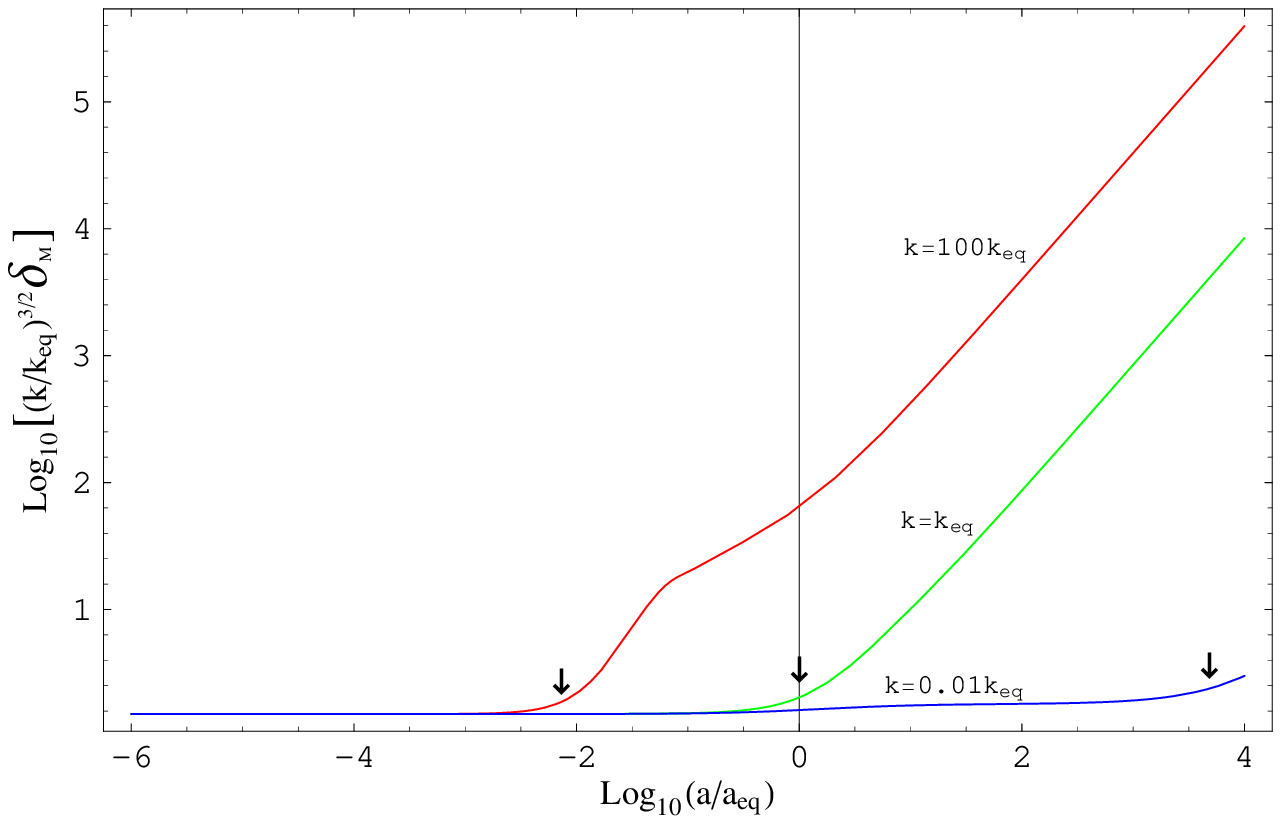}
\end{center}
\caption{Evolution of $|\delta_m|$ for 3 different modes. The modes are labelled  by their wave numbers
and the epochs at which they enter the Hubble radius are shown by small arrows. All the modes
remain frozen when they are outside the Hubble radius and grow linearly in the matter dominated
phase once they are inside the Hubble radius. The mode that enters the Hubble radius in 
the radiation dominated phase grows logarithmically until $y=y_{\rm eq}$.  These features are well
approximated by \eqs{deltam}, (\ref{eightsix}).}
\label{fig:dm}
\end{figure}

Given these initial conditions, it is fairly easy to integrate the equations forward in time and the numerical results are shown in Figs~\ref{fig:phi}, \ref{fig:dr1}, \ref{fig:dr}, \ref{fig:dm}. (In the figures $k_{eq}$ is taken to be $a_{eq}H_{eq}$.) To understand the nature of the evolution, it is, however, useful to try out a few analytic approximations to \eqs{first} -- (\ref{three}) which is what we will do now.

   \subsection{Evolution for $\lambda\gg d_H$}  
   
Let us begin by considering very large wavelength modes corresponding to the $k\eta\to 0$ limit. In this case adiabaticity is respected
and we can set $\delta_R\approx(4/3)\delta_m.$ Then  \eqs{first}, (\ref{sec}) become
   \begin{equation}
   y\Phi'+\Phi\approx -\frac{3y+4}{8(1+y)}\delta_R;\quad \delta_R'\approx 4\Phi'
   \end{equation}
 Differentiating the first equation and using the second to eliminate $\delta_m$, we get a second order equation for $\Phi$. Fortunately, this equation has an exact solution 
   \begin{equation}
   \Phi=\Phi_i\frac{1}{10 y^3}\left[16\sqrt{(1+y)}+9y^3+2y^2-8y-16\right]; \quad \delta_R\approx 4\Phi-6\Phi_i
   \label{exactsol}
   \end{equation}
[There is simple way of determining such an exact solution, which we will describe in Sec. \ref{sec:alt}.]. The initial condition on $\delta_R$ is chosen such that it goes to $-2\Phi_i$ initially.
The solution shows that, as long as the mode is bigger than the Hubble radius, the potential changes very little; it is constant initially as well as in the final matter dominated phase. At late times $(y\gg 1)$ we see that $\Phi\approx (9/10) \Phi_i$ so that  $\Phi$  decreases only by a factor (9/10) during the entire evolution if $k\to 0$ is a valid approximation.

\subsection{Evolution for $\lambda\ll d_H$ in the radiation dominated phase }\label{sec:evollambda}    
 
 When the mode enters Hubble radius in the radiation dominated phase, we can no longer ignore the pressure terms. The
 pressure makes radiation density contrast oscillate and the gravitational potential, driven by this, also oscillates  with a decay in the overall amplitude. An approximate procedure to describe this phase is to solve the coupled $\delta_R-\Phi$ system, ignoring $\delta_m$ which is sub-dominant and {\it then} determine $\delta_m$ using the form of $\Phi$.
 
When $\delta_m$ is ignored, the problem reduces to the one solved earlier in Eqs (\ref{sixtya}), (\ref{sixthree}) with $w=1/3$ giving
$\nu=3$.  Since $J_{3/2}$
 can be expressed in terms of trigonometric functions, the solution   given by \eq{sixtya} with $\nu = 3$, simplifies
to 
   \begin{equation}
   \Phi=\Phi_i\frac{3}{l^3y^3}\left[\sin(ly)-ly\cos(ly)\right];\quad l^2=\frac{k^2}{3k_c^2}
   \label{Phieqn}
   \end{equation}
Note that  as $y\to0$, we have $\Phi=\Phi_i,\Phi'=0$. This solution shows that once the mode enters the Hubble radius, the potential decays in an oscillatory manner. 
For $ly \gg 1$, the potential becomes $\Phi \approx -3\Phi_i (ly)^{-2} \cos(ly)$. In the same limit,
we get from \eq{sixthree} that 
\begin{equation}
\delta_R \approx-\frac{2}{3}k^2\eta^2\Phi\approx - 2l^2 y^2 \Phi \approx 6\Phi_i \cos(ly)
\label{rados}
\end{equation}
(This is analogous to \eq{six9} for the radiation dominated case.)
This oscillation is seen clearly in Fig~\ref{fig:dr1} and Fig.\ref{fig:dr} (left panel). The amplitude of oscillations is accurately captured by \eq{rados} for $k=100k_{eq}$ mode but not for $k=k_{eq}$; this is to be  expected since the mode is not entering in the radiation dominated phase. 
 
Let us next consider matter perturbations during this phase. They grow, driven by the gravitational potential determined above.  When $y\ll 1$, Eq.(\ref{sec}) becomes:
\begin{equation}
 \delta_m'' +  \frac{1}{y} \delta_m' = 3  \Phi'' + \frac{3}{y} \Phi' - 
   \frac{k^2}{k_c^2}  \Phi
   \label{eightyone}
\end{equation} 
The $\Phi$ is essentially determined by radiation and satisfies \eq{poteqn}; using this, we can rewrite \eq{eightyone} as
\begin{equation}
\frac{d\ }{dy}(y\delta_m') = -9(\Phi'+ \frac{2}{3}l^2y\Phi)
\label{eightytwo}
\end{equation}  
The general solution to the homogeneous part of \eq{eightytwo} (obtained by ignoring the right hand side) is $(c_1 + c_2 \ln y)$; hence the general solution to this equation is
\begin{equation}
\delta_m=(c_1 + c_2 \ln y)-9\int \frac{dy}{y}\int^y dy_1[\Phi'(y_1)+ \frac{2}{3}l^2y_1\Phi(y_1)]
\label{deltam}
\end{equation} 
For $y\ll1$ the  growing mode varies as  $\ln y$ and dominates over the rest; hence we conclude that,
 matter, driven by  $\Phi$, grows logarithmically during the radiation dominated phase for modes which are inside the Hubble radius.
 
 \subsection{Evolution in the matter dominated phase } 
  
 Finally let us consider the matter dominated phase, in which we can ignore the radiation and concentrate on \eq{first} and \eq{sec}. When $y\gg1$ these equations become:
 \begin{equation}
 y\Phi'+\Phi \approx -\frac{1}{2}\delta_m-\frac{k^2y}{3k_c^2}\Phi;\qquad
 y\delta_m''+\frac{3}{2}\delta_m'=-\frac{k^2}{k_c^2}\Phi
 \end{equation}
 These have a simple solution which we found earlier (see \eq{sixtynine}): 
 \begin{equation}
 \Phi=\Phi_\infty={\rm const.}; \qquad \delta_m=-2\Phi_\infty-\frac{2k^2}{3k_c^2}\Phi_\infty y \sim y
 \label{eightsix}
 \end{equation}
 In this limit, the matter perturbations grow linearly with expansion: $\delta_m\propto y\propto a$. In fact this is the most dominant growth mode in the linear perturbation theory.

   \subsection{An alternative description of matter-radiation system} \label{sec:alt} 

Before proceeding further, we will describe an alternative procedure for discussing the perturbations in dark matter and radiation, which has some advantages. In the formalism we used above, we used     
perturbations in the energy density of  radiation ($\delta_R$) and matter $(\delta_{m})$ as the dependent variables.  Instead, we now use perturbations in the \textit{total} energy density, $\delta$ and 
the perturbations in the entropy per particle, $\sigma$ as the new dependent variables. In terms of
$\delta_R,\delta_{m}$, these variables are defined as:

 \begin{equation}  
 \delta \equiv \frac{\delta\rho_{\rm total}}{\rho_{\rm total}} = \frac{\rho_R\delta_R + \rho_m  \delta_m}{\rho_R+\rho_m} = \frac{\delta_R+y\delta_m}{1+y}; \quad  y = \frac{\rho_m}{\rho_R} = \frac{a}{a_{\rm eq}}
 \end{equation}   

\begin{equation}  
\sigma  \equiv \left(\frac{\delta s}{s}\right) = \frac{3\delta T_R}{T_R} - \frac{\delta \rho_m}{\rho_m}= \frac{3}{4} \delta_R - \delta_m = \frac{\delta n_R}{n_R} - \frac{\delta n_m}{n_m}
\label{defofent}
\end{equation}   
Given the equations for $\delta_R,\delta_{m}$, one can obtain the corresponding equations for the new variables $(\delta, \sigma)$  by straight forward algebra. It is convenient to express them as two coupled equations for $\Phi$ and $\sigma$. After some direct but a bit tedious algebra, we get:
\begin{equation}  
y \Phi'' + \frac{y\Phi'}{2(1+y)} + 3 (1+ c_s^2) \Phi' + \frac{3 c_s^2\Phi}{4(1+y)} 
+c_s^2\frac{k^2}{k_c^2}\frac{y}{1+y}\Phi
= \frac{3 c_s^2\sigma}{2(1+y)}
\label{forphi}
\end{equation}
\begin{equation}  
y \sigma'' + \frac{y\sigma'}{2(1+y)} + 3c_s^2 \sigma' + \frac{3 c_s^2y^2}{4(1+y)}\frac{k^2}{k_c^2} \sigma
= \frac{ c_s^2y^3}{2(1+y)}\left(\frac{k}{k_c}\right)^4\Phi
\label{forsigma}
\end{equation}
where we have defined
  \begin{equation}  c_s^2 = \frac{(4/3) \rho_R}{4\rho_R + 3  \rho_m} = \frac{1}{3} \left( 1 + \frac{3}{4} \frac{\rho_m }{\rho_R}\right)^{-1} = \frac{1}{3} \left( 1 + \frac{3}{4} y \right)^{-1}
  \end{equation}  
These equations show that the entropy perturbations and gravitational potential (which is directly related to total energy density perturbations) act as sources for each other. The coupling between the two arises through the right hand sides of \eq{forphi} and \eq{forsigma}. We also see that if we set $\sigma=0$ as an initial condition, this is preserved to ${\cal O}(k^4)$ and --- for long wave length modes --- the $\Phi$ evolves independent of $\sigma$. The solutions to the coupled equations obtained by numerical integration is shown in Fig.(\ref{fig:phi}) right panel. The entropy perturbation $\sigma\approx0$  till the mode enters Hubble radius and grows afterwards tracking either $\delta_R$ or $\delta_m$ whichever is the dominant energy density perturbation.
To illustrate the behaviour of $\Phi$, let us consider the
adiabatic perturbations at large scales with $\sigma\approx0, k\to 0$; then the gravitational potential satisfies the equation:
\begin{equation}  
y \Phi'' + \frac{y\Phi'}{2(1+y)} + 3 (1+ c_s^2) \Phi' + \frac{3 c_s^2\Phi}{4(1+y)}
 = \frac{3 c_s^2\sigma}{2(1+y)}\approx 0
\end{equation}
which has the two independent solutions:
\begin{equation}  f_1(y) = 1 + \frac{2}{9y} - \frac{8}{9y^2} - \frac{16}{9 y^3}, \qquad
f_2(y) = \frac{\sqrt{1+y}}{y^2}
\end{equation} 
both of which diverge as $y\to 0$. 
We need to combine these two solutions  to find the  general solution, keeping in mind that the general solution should be
nonsingular and  become a constant (say, unity)  as $y\to0$. This fixes the linear combination uniquely:
\begin{equation}
f(y)=\frac{9}{10}f_1+\frac{8}{5}f_2=\frac{1}{10 y^3}\left[16\sqrt{(1+y)}+9y^3+2y^2-8y-16\right]
\end{equation}
Multiplying by $\Phi_i$ we get the solution that was found earlier (see Eq.~(\ref{exactsol})).
Given the form of $\Phi$ and  $\sigma \simeq 0$ we can determine all other quantities. In particular, we get:
 \begin{equation}  
 \delta_R= \frac{-2(1+y)d(y\Phi)/dy + y\sigma}{1+(3/4)y} \simeq - \frac{2 (1+y)}{1+(3/4)y} \frac{d}{dy} (y\Phi)
 \label{ninefive}
 \end{equation}
The corresponding velocity field, which we quote for future reference, is given by: 
\begin{equation}
 v_\alpha = - \frac{3c_s^2}{2(\dot a/a)} (1+y) \nabla_\alpha \frac{d(y\Phi)}{dy}
 \label{ninesix}
\end{equation}

We conclude this section by mentioning another useful result related to \eq{forphi}. 
   When $\sigma \approx 0$, 
   the equation for $\Phi$ can be re-expressed as 
   \begin{equation}
   a\frac{d\zeta}{da} = - \frac{2c_s^2}{3} \frac{k^2/a^2}{H^2} \frac{\rho}{\rho+p}\Phi\approx 0\quad ({\rm for}\ \frac{k}{aH} \ll 1)
   \label{zetacons}
   \end{equation}   
   where  we have defined:
   \begin{equation}
   \zeta = \frac{2}{3} \frac{\rho}{\rho+p} \frac{a}{\dot a} \left( \dot\Phi + \frac{\dot a}{a}\Phi\right) +\Phi =\frac{H}{\rho+p}\frac{ik^\alpha}{k^2}\delta T^0_\alpha +\Phi
   \label{zetaeqn}
   \end{equation}
   (The $i$ factor arises because of converting a gradient to the ${\bf k}$ space; of course, when everything is done correctly, all physical quantities will be real.) Other equivalent alternative forms for $\zeta$, which are useful are:
   \begin{equation}
   \zeta=\frac{2}{3[1+w(a)]}\frac{d}{da}(a\Phi) + \Phi=\frac{H^2}{a(\rho+p)}\frac{d}{dt}\left(\frac{a\Phi}{H}\right)
   \label{altform} 
   \end{equation}
For modes which are bigger than the Hubble radius, \eq{zetacons} shows that $\zeta$ is conserved.    
 When $\zeta$=constant, we can integrate \eq{altform} easily to obtain:
   \begin{equation}
   \Phi=c_1\frac{H}{a}+c_2\left[1-\frac{H}{a}\int_0^a\frac{da'}{H(a')}\right]
   \label{simpphi}
   \end{equation}
   This is the easiest way to obtain the solution in Eq.~(\ref{exactsol}).
   
   The conservation law for $\zeta$ also allows us to understand in a simple manner our previous result that $\Phi$ only deceases by a factor $(9/10)$ when the mode remains bigger than Hubble radius as we evolve the equations from $y\ll 1$ to $y\gg1$.   Let us compare the values of $\zeta$
 early in the radiation dominated phase and late in the matter dominated phase.
  From the first equation in \eq{altform}, [using $\Phi'\approx 0$] we find that,
  in the radiation dominated phase,
   $\zeta \approx (1/2)\Phi_i + \Phi_i = (3/2) \Phi_i$; late in the matter dominated phase, 
   $\zeta \approx (2/3)\Phi_f +\Phi_f =  (5/3) \Phi_f$. Hence the conservation of $\zeta $ gives
   $\Phi_f = (3/5)(3/2)\Phi_i = (9/10) \Phi_i$ which was the result obtained earlier. The expression in
\eq{simpphi} also works at late times in the $\Lambda$ dominated or curvature dominated universe.

One key feature which should be noted in the study of linear perturbation theory is the different amount of growths for $\Phi,\delta_R$ and $\delta_m$. The $\Phi$ either changes very little or decays; the $\delta_R$ grows in amplitude only by a factor of few. The physical reason, of course, is that the amplitude is frozen at super-Hubble scales and the pressure prevents the growth at sub-Hubble scales. In contrast, $\delta_m$, which is pressureless, grows logarithmically in the radiation dominated era and linearly during the matter dominated era. Since the later phase lasts for a factor of $10^4$ in expansion, we get a fair amount of growth in $\delta_m$.

 \section{ Transfer Function for matter perturbations}
 
 We   now have all the ingredients  to evolve the matter  perturbation from an initial value $\delta=\delta_i$ at $y=y_i\ll 1$ to the current epoch $y = y_0= a_{\rm eq}^{-1}$ in the matter dominated phase at $y\gg 1$. Initially, the wavelength of the perturbation will be bigger than the Hubble radius and the perturbation will essentially remain frozen.
 When it enters the Hubble radius in the radiation dominated phase, it begins to grow but only logarithmically (see section \ref{sec:evollambda} ) until the universe becomes matter dominated. In the final
 matter dominated phase, the perturbation grows linearly with expansion factor. The relation between
 final and initial perturbation can be obtained by combining these results. 
 
 Usually, one is more interested in the power spectrum $P_k(t) $ and the power per logarithmic band in $k-$space $\Delta_k(t)$. These quantities are defined in terms of 
 $\delta_k(t)$ through the equations:
 \begin{equation} 
 P_k(t)  \equiv 
  |\delta_k(t)|^2;\quad \Delta^2_k(t)\equiv{k^3 P_k(t)\over 2\pi^2}
   \end{equation}
 It is therefore convenient to study the evolution of $k^{3/2}\delta_k$ since its square will
 immediately give the  power per logarithmic band $\Delta_k^2$  in $k-$space. 
 
 Let us first consider a mode
 which enters the Hubble radius in the radiation dominated phase at the epoch $a_{\rm enter}$.
 From the scaling relation, 
  $a_{ent}/k\propto t_{ent}\propto a_{ent}^2$ we find that  $y_{ent}=(k_{eq}/k)$. Hence
 \begin{equation}
 k^{3/2}\delta_m(k,a=1)\  = \underbrace{\frac{1}{a_{eq}}}_{\rm MD}\  \underbrace{\ln\left(\frac{a_{eq}}{a_{ent}}\right)}_{\rm RD}\  \underbrace{\left[k^{3/2}\delta_{ent}(k)\right]}_{\rm at\ entry}
  \propto\  \ln\left(\frac{k}{k_{eq}}\right)[k^{3/2}\delta_{ent}(k)]
 \end{equation}
 where two factors --- as indicated --- gives the growth in radiation (RD) and matter dominated (MD) phases.
 Let us next consider the modes that enter in the matter dominated phase. In this case,
  $a_{ent}/k\propto t_{ent}\propto a_{ent}^{3/2}$ so that $y_{ent}=(k_{eq}/k)^2$. Hence
 \begin{equation}
 k^{3/2}\delta_m(k,a=1)= \underbrace{\frac{1}{a_{ent}}}_{\rm MD}\  \underbrace{\left[k^{3/2}\delta_{ent}(k)\right]}_{\rm at\ entry}
 \propto k^2[k^{3/2}\delta_{ent}(k)]
 \end{equation}
To proceed further, we need to know the $k-$dependence of the perturbation when it enters
the Hubble radius which, of course,  is related to the mechanism that generates the initial power spectrum.
The most natural choice will be that all the modes enter the Hubble radius with a constant amplitude
at the time of entry. This would imply that the physical perturbations are scale invariant
at the time of entering the Hubble radius, a possibility that was suggested by Zeldovich and Harrison \cite{zeldovich72} (years before inflation was invented!). We will see later that this is also true
for perturbations generated by inflation and thus is a reasonable assumption at least in
such models. Hence we shall assume
 \begin{equation} 
 k^{3}|\delta_{ent}(k)|^2=k^3P_{ent}(k)=C={\rm constant},
 \label{scaleinv}
  \end{equation}
  Using this we find that the current value of perturbation is given by
  \begin{equation}
  P(k, a=1)\propto \Big|\delta_m(k,a=1)\Big|^2\propto
 \begin{cases}
  k \text{ (for $k\ll k_{eq}$)}\\
  k^{-3}(\ln k)^2 \text{(for  $k\gg k_{eq}$)}
   \end{cases}
   \label{oneofive}
   \end{equation} 
   The corresponding power per logarithmic band is 
    \begin{equation}
  \Delta^2(k, a=1)\propto k^3\Big|\delta_m(k,a=1)\Big|^2\propto
 \begin{cases}
  k^4 \text{ (for $k\ll k_{eq}$)}\\
  (\ln k)^2 \text{(for  $k\gg k_{eq}$)}
   \end{cases}
   \end{equation}   
   The form for $P(k)$ shows that the evolution imprints the scale $k_{eq}$ on the power spectrum 
   even though the initial power spectrum is scale invariant. For $k<k_{eq}$ (for large spatial
   scales), the primordial form of the spectrum is preserved and the evolution only increases
   the amplitude preserving the shape. For $k>k_{eq}$ (for small spatial scales), the shape
   is distorted  and in general the power is suppressed in comparison with larger spatial scales.
   This arises because modes with small wavelengths enter the Hubble radius early on and have to wait
   till the universe becomes matter dominated in order to grow in amplitude. This is in contrast
   to modes with large wavelengths which continue to grow. It is this effect which suppresses
   the power at small wavelengths (for $k>k_{eq}$) relative to power at larger wavelengths.

\section{Temperature anisotropies of CMBR}\label{sec:tempcmbr}

We shall now apply the formalism we have developed to understand the temperature anisotropies in the 
cosmic microwave background radiation which is probably the most useful application of \textit{linear}
perturbation theory. We shall begin by developing the general formulation and the terminology
which is used to describe the temperature anisotropies.

Towards
every direction in the sky, \({\bf n}=( \theta,\psi)\) 
we can define  a fractional temperature fluctuation $\Delta({\bf n})\equiv (\Delta T/ T) (\theta,\psi)$.
Expanding this quantity in spherical harmonics on the sky plane as well as in terms of the 
spatial Fourier modes, we get the two relations:
\begin{equation} \Delta(\mathbf{n}) \equiv {\Delta T\over T} (\theta,\psi){ = }
 \sum_{l,m}^\infty a_{lm} Y_{lm}(\theta, \psi){=}
 \int \frac{d^3k}{(2\pi)} \Delta (\mathbf{k}) e^{i\mathbf{k}\cdot \mathbf{n} L} 
 \end{equation}
 where $L= \eta_0 - \eta_{_{\rm LSS}} $ is the distance to the last scattering surface (LSS) from which we are receiving the radiation.
  The last equality allows us to define the expansion coefficients $a_{lm}$ in terms of the temperature fluctuation
  in the Fourier space $\Delta (\mathbf{k})$ .   Standard identities of mathematical
  physics now give           
  \begin{equation}
  a_{lm} = \int \frac{d^3k}{(2\pi)^3}\  (4\pi)\  i^l\  \Delta (\mathbf{k})\  j_l(kL)\  Y_{lm}({\hat {\bf k}})
  \label{alm}
  \end{equation}
  Next, let us consider the angular correlation function of temperature anisotropy, which is given by:  
 \begin{equation} {\cal C} (\alpha) = \langle \Delta(\bld n) \Delta(\bld m)\rangle = \sum\sum \langle a_{lm}a_{l'm'}^*\rangle Y_{lm}(\bld n)Y_{l'm'}^*(\bld m).
 \label{oneonine}
 \end{equation}
 where the wedges denote an ensemble average. For a Gaussian random field of fluctuations we can 
 express the ensemble average as
 $  \langle a_{lm}a_{l'm'}^*\rangle = C_l \delta _{ll'} \delta_{mm'}$. Using  \eq{alm}, we get a
 relation between $C_l$ and $\Delta(k)$.
   Given \( \Delta(k)\), the  \(C_l\)'s are given by:
  \begin{equation} C_l = \frac{2}{\pi} \int_0^\infty k^2 dk\  |\Delta (k)|^2 \ j^2_l(kL)
  \label{cleqn}
  \end{equation}
 Further, \eq{oneonine} now becomes:
  \begin{equation} {\cal C} (\alpha) = \sum_l {(2l+1)\over 4\pi} C_l P_l(\cos \alpha)
  \label{oneten}
  \end{equation} 
  Equation~(\ref{oneten})  shows that the mean-square value of temperature fluctuations and the quadrupole anisotropy
corresponding to $l=2$ are given by
\begin{equation} \left( {\Delta T\over T}\right)^2_{\rm rms} = {\cal C}(0) = {1\over 4\pi} \sum_{l=2}^\infty \left( 2l+1 \right) C_l, 
\quad \left( {\Delta T\over T}\right)^2_Q = {5\over 4\pi} C_2.
\label{oneeleven}
\end{equation}  
These can be explicitly computed if we know $\Delta(k)$ from the perturbation theory. 
(The motion of our local group through the CMBR leads to a large $l=1$ dipole contribution
in the temperature anisotropy. In the analysis of CMBR anisotropies, this is usually
subtracted out. Hence the leading term is the quadrupole with $l=2$.)
    
   It should be noted that, for
   a given $l$, the $C_l $ is the average over all $m = -l, ...-1,  0, 1, ... l$. 
   For a Gaussian random field, one can also compute the variance around this mean value.
   It can be shown that this  variance in $C_l$ is $2 C_l^2/(2l + 1)$. 
   In other words, there is an intrinsic root-mean-square fluctuation in the observed, mean value of
   $C_l$'s which is of the order of $\Delta C_l/C_l \approx (2l+1)^{-1/2}$.
   It is not possible for any CMBR observations which measures the $C_l$'s to reduce
   its uncertainty below this intrinsic variance --- usually called 
    the ``cosmic variance''.
   For large values of $l$, the cosmic variance is usually sub-dominant to other 
   observational errors but for low $l$ this is the dominant source of uncertainty
   in the measurement of $C_l$'s. Current WMAP observations are indeed only limited by cosmic variance at low-$l$.
   
   As an illustration of the formalism developed above, let us compute the $C_l$'s
   for low $l$ which will be contributed essentially by fluctuations at large
   spatial scales. Since these fluctuations will be dominated by gravitational
   effects, we can ignore the complications arising from baryonic physics and compute
   these using the formalism we have developed earlier. 

   We begin by noting that the   
 redshift law of photons in the unperturbed Friedmann universe, $\nu_0=\nu(a)/a$, gets modified
 to the form 
 $\nu_0 = \nu(a)/[a(1+\Phi)]$ in a perturbed FRW universe. The argument of the Planck
 spectrum will thus scale as   
   \begin{equation}
   \frac{\nu_0}{T_0}=\frac{\nu(a)}{aT_0(1+\Phi)} =\frac{\nu(a)}{a\langle T_0\rangle[1+(\delta_R/4)](1+\Phi)}
   \cong\  \frac{\nu(a)}{a\langle T_0 \rangle[1+\Phi+(\delta_R/4)]}
   \end{equation} 
 This is equivalent to a temperature fluctuation of the amount
 \begin{equation}  
 \left(\frac{\Delta T}{T}\right)_{\rm obs} = \frac{1}{4}\delta_R + \Phi
 \end{equation}
at large scales. 
(Note that the observed $\Delta T/T$ is not just $(\delta_R/4)$ as one might have naively imagined.)
To proceed further, we recall our large scale solution (see \eq{exactsol}) for the gravitational potential:
   \begin{equation}
   \Phi=\Phi_i\frac{1}{10 y^3}\left[16\sqrt{(1+y)}+9y^3+2y^2-8y-16\right];
   \quad\delta_R=4\Phi-6\Phi_i
   \end{equation}
   At $y=y_{dec}$ we can take the asymptotic solution $\Phi_{dec}\approx (9/10)\Phi_i$.
    Hence we get 
   \begin{equation}
 \left(\frac{\Delta T}{T}\right)_{\rm obs} =  \left[\frac{1}{4}\delta_R+\Phi\right]_{dec}=2\Phi_{dec}-\frac{3}{2}\Phi_i\approx2\Phi_{dec}-\frac{3}{2}\frac{10}{9}\Phi_{dec}=\frac{1}{3}\Phi_{dec}
 \label{phiby3}
   \end{equation}
 We thus obtain the nice result that the observed temperature fluctuations at very large scales
 is simply related to the fluctuations of the gravitational potential at these scales. (For a discussion of
 the $1/3$ factor, see \cite{hwang}).
 Fourier transforming this result we get $ \Delta (\mathbf{k}) = (1/3) \Phi (\mathbf{k},\eta_{_{\rm LSS}})$ where $\eta_{_{LSS}}$ is the  conformal time at  the last scattering surface. 
 (This contribution is called Sachs-Wolfe effect.)
 It follows from \eq{cleqn} that the contribution to $C_l$ from the gravitational potential is 
\begin{equation}
C_l =\frac{2}{\pi} \int k^2 dk |\Delta(k)|^2 j_l^2(kL) = \frac{2}{\pi} \int_0^\infty \frac{dk}{k} 
\frac{k^3|\Phi_k|^2}{9} j_l^2 (kL)
\label{clgrav}
\end{equation}
with 
\begin{equation}
L=\eta_0 - \eta_{_{\rm LSS}} \approx \eta_0 \approx 2 (\Omega_m H_0^2)^{-1/2} \approx 6000\ \Omega_m^{-1/2}\ h^{-1}\ {\rm Mpc}
\end{equation}
For a scale invariant spectrum, $k^3|\Phi_k|^2$ is a constant independent of $k$. (Earlier on, in \eq{scaleinv} we said that scale invariant spectrum has
$k^3|\delta_k|^2=$ constant. These statements are equivalent since $\delta\approx -2\Phi$ at the large scales because of \eq{eightsix} with the extra correction term in \eq{eightsix}  being about $3\times 10^{-4}$ for $k\approx L^{-1},y=y_{dec}$.) As we shall see later, inflation generates such a perturbation. In this case, it is conventional to introduce a constant amplitude $A$ and write:
 \begin{equation} 
 \Delta_\Phi^2 \equiv 
{k^3|\Phi_k|^2\over 2\pi^2} = A^2 = {\rm constant}
\label{spect}
\end{equation}
  Substituting this form into \eq{clgrav} and evaluating the integral, we find 
  that 
\begin{equation} 
\frac{l(l+1)C_l}{2\pi} = \left( \frac{A}{3} \right)^2
\label{ctwo}
\end{equation}  
As an application of this result, let us consider the observations of COBE which measured
the temperature fluctuations for the first time in 1992. This satellite obtained
the RMS fluctuations and the quadrupole after smoothing over an angular scale of 
about $\theta_c \approx 10^\circ$. Hence the observed values are slightly different from those in \eq{oneeleven}. We have, instead,
  \begin{equation} \left({\Delta T\over T}\right)^2_{{\rm rms}}=
{1\over 4\pi}
\sum^{\infty}_{l=2}(2l+1)
C_l \exp\left(-{l^2\theta^2_c\over 2}\right); \quad
   \left({\Delta T\over T}\right)^2_Q={5\over 4\pi}C_2
e^{-2\theta^2_c}.
\end{equation}
Using \eqs{spect}, (\ref{ctwo}) we find that 
  \begin{equation} \left({\Delta T\over T}\right)_Q\cong
0.22A;\quad
\left({\Delta T\over T}\right)_{{\rm rms}}\cong 0.51A.
\end{equation}
Given these two measurements, one can verify that the fluctuations are consistent with the 
scale invariant spectrum by checking their ratio. Further, the numerical
value of the observed $(\Delta T/T)$ can be used to determine the amplitude $A$. One finds that
$A\approx 3\times 10^{-5} $ which sets the scale of fluctuations in the gravitational potential
at the time when the perturbation enters the Hubble radius.

Incidentally, note that the solution $\delta_R=4\Phi-6\Phi_i$ corresponds to $\delta_m=(3/4)\delta_R=3\Phi-(9/2)\Phi_i$. At $y=y_{dec}$, taking $\Phi_{dec}=(9/10)\Phi_i$, we get
$\delta_m=3\Phi_{dec}-(9/2)(10/9)\Phi_{dec}=-2\Phi_{dec}$. Since $(\Delta T/T)_{obs}=(1/3)\Phi_{dec}$
 we get $\delta_m=-6(\Delta T/T)_{obs}$. This shows that the amplitude of matter perturbations is a factor six larger that the amplitude of temperature anisotropy for our adiabatic initial conditions. In several other models, one gets $\delta_m=\mathcal{O}(1)(\Delta T/T)_{obs}$. So, to reach a given level of nonlinearity in the matter distribution at later times, these models will require higher values of
$(\Delta T/T)_{obs}$ at decoupling. This is one reason for such models to be observationally ruled out.

There is another useful result
which we can obtain from \eq{cleqn} along the same lines as we derived the Sachs-Wolfe effect.
Whenever $k^3|\Delta(k)|^2$ is a slowly varying function of $k$, we can pull out this factor
out of the integral and evaluate the integral over $j_l^2$. This will give the result for any
slowly varying $k^3 |\Delta(k)|^2$
\begin{equation}
\frac{l(l+1)C_l}{2\pi} \approx  \left( \frac{k^3 |\Delta(k)|^2}{2\pi^2} \right)_{kL\approx l}
\end{equation}    
This is applicable even when different processes contribute to temperature anisotropies 
as long as they add in quadrature. While far from accurate, it allows one to estimate the
effects rapidly.

 \subsection{ CMBR Temperature Anisotropy: More detailed theory}  
 
 We shall now work out a more detailed theory of temperature anisotropies of CMBR so that one
  can understand the effects at small scales as well. A convenient starting point is the 
 distribution function for photons with perturbed Planckian distribution, which we can write as: 
  \begin{equation}  f(x^\alpha, \eta, E, n^\alpha) = \frac{ I_\nu}{2 \pi  \nu^3 } = f_P \left(\frac{aE}{1 + \Delta}\right); \qquad f_P(\epsilon) \equiv 2 \left[ \exp \left( \epsilon/T_0\right) - 1 \right]^{-1}\end{equation} 
  The $f_p(\epsilon)$ is the standard Planck spectrum for energy $\epsilon$ and we take $\epsilon = a E ( 1+ \Delta)^{-1}$ to take care of the perturbations.
  In the absence of collisions, the distribution function is conserved along the trajectories of photons so that   $df/d\eta =0$. So, in the presence of collisions, we can write the 
 time evolution of the distribution function  as
\begin{equation}  \frac{df}{d\eta} = \left(\frac{aE}{1 + \Delta} \right)\, f_P'\left(\frac{aE}{1 + \Delta}\right)\left[ \frac{d \ln (aE)}{d\eta} - \frac{d\Delta}{d\eta} \right] = \left(\frac{df}{d\eta}\right)_{\rm coll}
\end{equation}  
where the right hand side gives the contribution due to collisional terms.  
Equivalently, in terms of $\Delta$, the same equation takes the form:
\begin{equation}  
 \frac{d\Delta}{d\eta} - \frac{d \ln (aE)}{d\eta} = -\left( \frac{1 + \Delta}{aE}\right) [f'_P]^{-1}\left(\frac{df}{d\eta}\right)_{coll} \equiv 
 \left(\frac{d\Delta}{d\eta}\right)_{\rm coll}
 \label{onetwoseven}
\end{equation} 
To proceed further, we need the expressions for the two terms on the left hand side. First term, on using the standard expansion for total derivative, gives: 
\begin{equation}  \frac{d\Delta}{d\eta} = \pdov{\Delta}{\eta} + \pdov{\Delta}{x^\alpha} \frac{dx^\alpha}{\underbrace{d\eta}_{n^\alpha}} + \frac{\partial\Delta}{\underbrace{\partial E}_{\rm zero}} \frac{dE}{d\eta} + \underbrace{\frac{\partial\Delta}{\partial n^\alpha} \frac{dn^\alpha}{d\eta}}_{\mathcal{O}(\Delta^2)=0} \cong \partial_\eta \Delta + n^\alpha \partial_\alpha \Delta 
\end{equation} 
(Note that we are assuming $\partial\Delta/\partial E=0$ so that the perturbations do not depend on the
frequency of the photon.) 
To determine the second term, we note that it vanishes in the unperturbed Friedmann universe and arises essentially due to the variation of $\Phi$. Both the intrinsic time variation of $\Phi$ as well as its variation along the photon path will contribute, giving:
\begin{equation}  \frac{d\ln(aE)}{d\eta} = - n^\alpha \partial_\alpha \Phi + \partial_\eta\Phi\end{equation} 
(The minus sign arises from the fact that the we have $(1+2\Phi)$ in $g_{00}$ but $(1-2\Phi)$ in the spatial perturbations.)
Putting all these together, we can bring the evolution  equation \eq{onetwoseven} to the form:
\begin{equation}  \frac{d\Delta}{d\eta}  = - n^\alpha \partial_\alpha \Phi + \partial_\eta \Phi + \left(\frac{d\Delta}{d\eta}\right)_{\rm coll}
\label{evoleqn}
\end{equation}  
Let us next consider the collision term, which can be expressed in the form:
\begin{eqnarray}
\left(\frac{d\Delta}{ad\eta}\right)_{\rm coll} &&= -N_e \sigma_{_T}  \Delta+N_e \sigma_{_T}\left( \frac{1}{4} \delta_R\right)+N_e \sigma_{_T} ( {\bf v} {\bf \cdot  n})\nonumber\\
&&=N_e \sigma_{_T} \left( - \Delta + \frac{1}{4} \delta_R + {\bf v} {\bf \cdot  n}
\right) 
\label{onethreeone}
\end{eqnarray}
Each of the terms in the right hand side of the  first line has a simple interpretation. The first term describes the removal
of photons from the beam due to Thomson scattering with the electrons while the second term gives the scattering contribution into the beam. In a static universe, we expect these two terms to cancel if $\Delta=(1/4)\delta_R$ which fixes the relative coefficients of these two terms. The third term is a correction due to the fact that the electrons which are scattering the photons are not  at rest relative to the cosmic frame. This leads to a Doppler shift which is accounted for by the third term. (We denote electron number \textit{density} by $N_e$ rather than $n_e$ to avoid notational conflict with $n^\alpha$.)

Formally, \eq{evoleqn} is a first order linear differential equation for $\Delta$. To eliminate the $-N_e\sigma_T \Delta$  term which is linear in $\Delta$ in the right hand side, we use the standard integrating factor $\exp(-\tau)$ where
 \begin{equation}  \tau (\chi) \equiv \int_0^\chi d\eta\, \left( a N_e \sigma_{_T}\right) 
 \label{onethreethree}
 \end{equation}  
We can then formally integrate \eq{evoleqn} to get:
\begin{equation}   \Delta  ( {\bf n}) = \int_0^{\eta_0} d\chi e^{-\tau (\chi)} \left[- n^\alpha \partial_\alpha \Phi + \partial_\eta \Phi + aN_e \sigma_{_T} \left(  \frac{1}{4} \delta_R +{\bf v} {\bf \cdot  n} \right)\right] 
\label{deltaofn}
\end{equation} 
We can write
\begin{equation}
e^{-\tau} (- n^\alpha \partial_\alpha \Phi) = -\left( \frac{d\Phi}{d\eta}\right) e^{-\tau} + (\partial_\eta \Phi)e^{-\tau} = - \frac{d}{d\eta} ( \Phi e^{-\tau}) + ( a N_e \sigma_{_T} \Phi) e^{-\tau} + ( \partial_\eta \Phi) e^{-\tau}
\end{equation}
On integration, the first term gives zero at the lower limit and an unimportant constant (which does not depend on $\mathbf{n}$). Using the rest of the terms,
we can write \eq{deltaofn} in the form:
\begin{eqnarray}
 \Delta  (\textbf{n}) &=& \int_0^{\eta_0} d\chi \, e^{-\tau} \left[ 
2 \partial_\eta \Phi + a N_e \sigma_{_T} 
\left( \Phi + \frac{1}{4} \delta_R + {\bf v} {\bf \cdot  n}\right)
\right]\nonumber\\
&=&\int_0^{\eta_0} d\chi \, e^{-\tau} \left[ 2 \partial_\eta \Phi\right]
 + \int_0^{\eta_0} d\chi \, (e^{-\tau}a N_e \sigma_{_T})
  \left( \Phi + \frac{1}{4} \delta_R + {\bf v} {\bf \cdot  n}\right)
  \label{onethreeoneeqn}
\end{eqnarray}
The first term gives the contribution due to the intrinsic time variation of the gravitational potential along the path of the photon and is called the integrated Sachs-Wolfe effect. In the second term one can make further simplifications. Note that 
 $e^{-\tau}$ is essentially unity (optically thin) for $z< z_{\rm rec}$ and zero (optically thick) for $z> z_{\rm rec}$; on the other hand, $N_e \sigma_T$ is zero for
$z<z_{\rm rec}$ (all the free electrons have disappeared) and is large for $z> z_{\rm rec}$. Hence the product \( (a N_e e^{-\tau} )\) is sharply peaked at $\chi = \chi_{\rm rec} $ (i.e. at \( z\simeq 10^3 \) with \( \Delta z \simeq 80 \)).
Treating this sharply peaked quantity as essentially a Dirac delta function (usually called the
instantaneous recombination approximation) we can approximate the second term in \eq{onethreeoneeqn} as a contribution occurring just on the LSS: 
\begin{equation}
 \Delta  (\textbf{n})  = \left( \frac{1}{4} \delta_R + {\bf v \cdot n} + \Phi\right)_{\rm LSS} + 2 \int_{\eta_{_{\rm LSS}}}^{\eta_0} d\chi \partial_\eta \Phi 
\label{onethreeseven}
\end{equation}  
In the second term we have put $\tau=\infty$ for $\eta < \eta_{\rm LSS}$ and $\tau=0$ for $\eta > \eta_{\rm LSS}$.

Once we know $\delta_R, \Phi$ and ${\bf v}$ on the LSS from perturbation theory, we can take
a Fourier transform of this result to obtain $\Delta(k)$ and use \eq{cleqn} to compute $C_l$.
At very large scales the velocity term is sub-dominant and we get back the Sachs-Wolfe effect
derived earlier  in \eq{spect}. For understanding the small scale effects, we need to introduce baryons into the picture which is our next task.  

\subsection{Description of photon-baryon fluid}  
  
To study the interaction of photons and baryons in the fluid limit, we need to again start from the continuity equation and Euler equation.  
  In Fourier space, the continuity equation is same as the one we had  before (see \eq{fiftyone}): 
  \begin{equation} \left(\frac{3}{4}\right) \dot\delta_R  = - k^2 v_R + 3 \dot \Phi; \qquad \dot \delta_B = - k^2 v_B + 3 \dot \Phi
  \label{onefourfive}
  \end{equation}  
The Euler equations, however, gets modified; for photons, it becomes:  
\begin{equation}  \dot v_R = (\frac{1}{4} \delta_R +\Phi) { - \dot \tau (v_R - v_B)}; \qquad \dot \tau  = N_e \sigma_{_T} a
\label{onefoursix}
\end{equation}
The first two terms in the right hand side are exactly the same as the ones in  \eq{fiftythree}. The last term is analogous to a viscous drag force between the photons and baryons which arises because of the non zero relative velocity between the two fluids. The coupling is essentially due to Thomson scattering which leads to the factor $\dot \tau$. (The notation, and the physics, is the same as in \eq{onethreethree}).
The corresponding Euler equation for the baryons is: 
\begin{equation}  \dot v_B= - \frac{\dot a}{a} \, v_B +  \Phi { + \frac{\dot \tau  (v_R - v_B)}{R}}
\label{dotvb}
\end{equation}
where
\begin{equation}  R \equiv \frac{p_B +\rho_B}{p_R+\rho_R} \simeq \frac{3\rho_B}{4\rho_R}\approx  30\  \Omega_Bh^2  \left(\frac{a}{10^{-3}}\right)\end{equation}  
Again, the first two terms in the right hand side of \eq{dotvb} are the same as what we had before in \eq{fiftythree}. The last term has the same interpretation as in the case of Euler equation \eq{onefoursix} for photons, except for the factor $R$. This quantity essentially takes care of the inertia of baryons relative to photons. Note that the 
 the conserved momentum density of photon-baryon fluid has the form
\begin{equation}
(\rho_R + p_R ) v_R + (\rho_B + p_B) v_B \approx (1+R) (\rho_R + p_R) v_R
\end{equation}
which accounts for the extra factor $R$ in \eq{dotvb}.

We can now combine the  \eqs{onefourfive}, (\ref{onefoursix}), (\ref{dotvb}) to obtain, 
to lowest order in \( ( k/\dot \tau ) \) the equation:
\begin{equation}  \ddot \delta_R + \frac{\dot R}{(1+ R) } \,\dot \delta_R + k^2 c_s^2 \delta_R = F\end{equation} 
with
\begin{equation}  F = 4\left[ \ddot \Phi + \frac{\dot R}{(1+R)} \dot \Phi - \frac{1}{3} k^2 \Phi\right]; \quad c_s^2 = \frac{1}{3(1+R)}\end{equation} 
An exact solution to this equation is difficult to obtain. 
However, we can try to understand several features by an approximate method in which 
we treat the time variation of $R$ to be small. In that case, we can drop the $\dot R$
terms on both sides of the equation. Since we know that the physically relevant temperature fluctuation is 
$\Delta=(1/4)\delta_R+\Phi$, we can recast the above equation for $\Delta$ as:
\begin{equation}
\ddot\Delta+k^2c_s^2\Delta\approx-k^2c_s^2R\Phi +2\ddot\Phi
\end{equation} 
Let us further ignore the time variation of all terms (especially $\ddot\Phi$ on the right hand side). Then, 
the solution is just $\Delta=-R\Phi+ A \cos ( kc_s \eta_{_{\rm LSS}}) + B\sin ( kc_s \eta_{_{\rm LSS}})$. To fix the initial conditions which determine $A$ and $B$, we recall that early on $(\eta\to0)$, we have $\Delta\to\Phi/3$ (see \eq{phiby3}) and corresponding velocity should vanish. This gives the solution:
\begin{equation} \frac{1}{4} \delta_R + \Phi = \frac{\Phi_i}{3} \, \left( 1 + 3 R\right) \, \cos ( kc_s \eta_{_{\rm LSS}}) - \Phi_i R; \qquad v = - \Phi_i\left( 1 + 3 R\right) \, c_s\sin ( kc_s \eta_{_{\rm LSS}}) \end{equation} 
(One can do a little better by using WKB approximation in which $( kc_s \eta_{_{\rm LSS}})$ can be replaced by the integral of $kc_s$ over $\eta$ but it is not very important.)
Given this solution, one can proceed as before and compute the $C_l$'s. 
Adding the effects of [$\Phi + (1/4)\delta_R$] and that of [${\bf v \cdot n}$] in quadrature
and noticing that the angular average of $\left<({\bf v \cdot n})^2\right> = (1/3) v^2$
we can estimate the $C_l$ 
for scale invariant ( \( k^{3}|\Phi_k|^2 = 2\pi^2 A^2 \)) spectrum to be:
\begin{equation}  l(l+1)C_l = 2\pi^2A^2\left\{ \left[ \frac{(1+3R)}{3} \cos(k^* c_s \eta_{_{\rm LSS}}) - R\right]^2 + \frac{(1+3R)^2}{3}  c_s^2\, \sin^2(k^* c_s \eta_{_{\rm LSS}}) \right\}
\end{equation}  
with $k^*L\approx l$ with $ L = \eta_0 - \eta_{_{\rm LSS}} \simeq \eta_0$.
The key feature is, of course, the maxima and minima which arises from the trigonometric functions. The peaks of $C_l$ are determined by the condition $ k^*c_s \eta_{_{\rm LSS}} = lc_s \eta_{_{\rm LSS}}/\eta_0=n\pi $; that is 
\begin{equation}
 l_{\rm peak}  = \frac{n\pi} {c_s}\left( \frac{\eta_0}{\eta_{_{\rm LSS}}}\right) 
=n\pi \sqrt{3} (1+z_{\rm dec})^{1/2} \approx 172n
\end{equation}
More precise work gives the first peak at \( l_{\rm peak} \simeq 200.\)
It is also clear that because of non zero $R$ the peaks are larger when the cosine term is 
negative; that is, the odd peaks corresponding to $n= 1, 3, ...$ have larger amplitudes
than the even peaks with $n= 2, 4, ...$.

Incidentally, the above approximation is not very good for modes which enter the Hubble radius
during the radiation dominated phase since $\Phi$ does evolve with time (and decays) in the 
radiation dominated phase. We saw that $\Phi \approx -3\Phi_i (ly)^{-2} \cos (ly)$ asymptotically
in this phase (see \eq{rados}).  From \eq{rados} we find that during this phase,
for modes which are inside the Hubble radius, we can take $\delta_R \approx 6\Phi_i\cos(ly)$, so that
$\Delta\approx\delta_R/4\approx (3/2)\Phi_i\cos(ly)$. On the other hand, at very large scale, the amplitude was
$\Delta=\Phi/3=(1/3)(9/10)\Phi_i=(3/10)\Phi_i$. Hence the amplitude of the modes that enter the horizon during the radiation dominated phase
is enhanced by a factor $(3/2)(10/9)=5$, relative to the large scale amplitude contributed by modes which enter during matter dominated phase.
This is essentially due to the driving term $\ddot\Phi$ which is nonzero in the radiation dominated phase but zero in the matter dominated phase.
(In reality, the enhancement is smaller because the relevant modes have $k\gtrsim k_{\rm eq}$ rather than $k\gg k_{\rm eq}$; see \figs{fig:dr1} and \ref{fig:dr}.)

If this were the whole story, we will see a series of peaks and troughs in the temperature anisotropies as a function of angular scale. In reality, however, there are processes which damp out the anisotropies at small angular scales (large -$l$) so that only the first few peaks and troughs are really relevant. We will now discuss two key damping mechanisms which are responsible for this.

The first one is the finite width of the last scattering surface which makes it uncertain from which event we are receiving the photons. 
In general, if ${\cal P}(z)$ is the probability that the photon was last scattered at redshift $z$, then we can write:
\begin{equation}
 \lb {\Delta T\over T}\rb_{\rm obs} = \int dz \ \left\{ 
 \begin{array}{ll}  &(\Delta T/T) \ \textrm {if the last} \\
& \textrm{scattering was at \(z\)}
\end{array}
\right\} \times  {\cal P}(z).
\end{equation}
From \eq{photonscat} we know that ${\cal P}(z)$ is a Gaussian with width $\Delta z=80$. This corresponds to a length scale
\begin{equation} \Delta l = c\lb {dt\over dz}\rb \Delta z \cdot (1+z_{\rm dec}) 
\approx H_0^{-1} {\Delta z \over \Omega^{1/2} z^{3/2}_{\rm dec}}
{\approx}
 8 \lb \Omega h^2 \rb^{-1/2} \, {\rm Mpc}.
\end{equation}
over which the temperature fluctuations will be smoothed out.

It turns out that there is another effect, which is slightly more important. This arises from the fact that the photon-baryon fluid is not tightly coupled and the photons can diffuse through the fluid. This diffusion can be modeled as a random walk and the root mean square distance traveled by the photon during this diffusion process will smear the temperature anisotropies over that length scale.    
This photon diffusion length scale can be estimated as follows:
 \begin{equation} 
 (\Delta x)^2= \underbrace{N}_{number of collisions}\underbrace{\left({q\over a}\right)^2}_{comoving meanfree path}
={\Delta t\over q(t)} {q^2\over a^2}    
={\Delta t\over a^2} q(t)            
 \end{equation}
Integrating, we find the mean square distance traveled by the photon to be 
 \begin{equation}
 x^2\equiv\int^{t_{\rm dec}}_0{dt\over a^2(t)} q(t)
={3\over 5}{t_{\rm dec}q(t_{\rm dec})\over a^2(t_{\rm dec})}  
\end{equation}  
The corresponding proper length scale below which photon diffusion will wipe out temperature anisotropies is:                  
  \begin{equation} q_{\rm diff}=a(t_{\rm dec}) x= \left[{3\over 5} t_{\rm dec} q(t_{\rm dec})
\right]^{\fra{1}{2}}\simeq
35\  {\rm Mpc} 
\left({\Omega_Bh^2\over 0.02}\right)^{-\fra{1}{2}}\left(\Omega h_{50}^2\right)^{-\fra{1}{ 4}}.
 \end{equation}
It turns out that this is the dominant sources of damping of temperature anisotropies at large $l\approx 10^3$.

\section{ Generation of initial perturbations from inflation }\label{sec:inflation}

In the description of linear perturbation theory given above, we assumed that some small perturbations existed in the early universe which are amplified through gravitational instability. To provide a complete picture we need a mechanism for generation of these initial perturbations.
One such mechanism is provided by inflationary scenario which allows for the quantum fluctuations in the field driving the inflation to provide classical energy density perturbations at a late epoch. (Originally inflationary scenarios were suggested as pseudo-solutions to certain pseudo-problems; that is only of historical interest today and the only reason to take the possibility of  an inflationary phase in the early universe seriously is  because it provides a mechanism for generating these perturbations.) We shall now discuss how this can come about.

The basic assumption in inflationary scenario is that the universe underwent a rapid --- nearly exponential --- expansion for a brief period of time in very early universe.  The simplest way of realizing such a phase is to postulate the existence of a scalar field with a nearly flat potential. The dynamics of the universe, driven by a
  scalar field source, is described by:
  \begin{equation}  \fra{1}{a^2}\left(\frac{da}{dt}\right)^2 = H^2(t)  = \fra{1}{3 M^2_{\rm Pl}}  \left[ V(\phi) + \fra{1}{2} \left(\frac{d\phi}{dt}\right)^2\right]; \qquad \frac{d^2 \phi}{dt^2} + 3H\frac{d\phi}{dt} = - \fra{dV}{d\phi}
   \label{onefivefive}
   \end{equation}    
  where $M_{pl} = (8\pi G)^{-1/2}$.
If the potential is nearly flat for certain range of $\phi$, we can introduce the      
 `` slow roll-over'' approximation, under which these equations become:
  \begin{equation}  H^2 \simeq \fra{V(\phi)}{3M_{\rm Pl}^2}; \qquad 3 H  \frac{d\phi}{dt} \simeq - V'(\phi)\end{equation}
For this slow roll-over to last for reasonable length of time, we need to assume that the terms ignored in the \eq{onefivefive} are indeed small. This can be quantified in terms of    the parameters:
\begin{equation}  \epsilon (\phi) = \frac{M_{\rm Pl}^2}{2} \left(\frac{V'}{V}\right)^2; \qquad \eta(\phi) =  M_{\rm Pl}^2\, \frac{V''}{V} \end{equation}  
 which are taken to be small. Typically the inflation ends when this assumption breaks down.
 If such an inflationary  phase lasts up to some time $t_{end}$ then the universe would have undergone an expansion by a factor $\exp N(t)$ during the interval $(t,t_{end})$ where
  \begin{equation}  N \equiv  \ln \fra{a(t_{\rm end})}{a(t)} = \int_t^{t_{\rm end}} H\, dt \simeq \fra{1}{M_{\rm Pl}^2} \int_{\phi_{\rm end}}^\phi \fra{V}{V'}\, d\phi
\end{equation} 
One usually takes $N\simeq 65$ or so.
 
 Before proceeding further, we would like to make couple of comments regarding such an inflationary phase. To begin with, it is not difficult to obtain \textit{exact} solutions for $a(t)$ with rapid expansion by tailoring the potential for the scalar field. In fact, given any $a(t)$ and thus a $H(t)=(\dot a/a)$, one can determine a potential  $V(\phi)$ for a scalar field such that \eq{onefivefive} are satisfied
 (see the first reference in \cite{tptachyon}). One can verify that, this is done by the choice: 
  \begin{equation} 
   V(t) = \frac{1}{16\pi G} H \left[6H + \frac{2}{H}\frac{dH}{dt} \right];\quad 
   \phi (t) = \int dt \left[ \frac{-2}{8\pi G}\frac{dH}{dt}\right]^{1/2} 
   \label{vteqn}
\end{equation}
Given any $H(t)$, these equations give $(\phi(t), V(t))$  and thus implicitly determine the necessary $V(\phi)$. As an example, note that
a power law inflation, $a(t) = a_0t^p$  (with $p\gg 1$) is generated by:
\begin{equation}  V(\phi) = V_0 \exp \left( - \sqrt{ \frac{2}{p}} \, \frac{\phi}{M_{\rm Pl}}\right)\end{equation}    
while an exponential of power law 
\begin{equation}  a(t) \propto \exp(At^f),\qquad f= \frac{\beta}{4+\beta}, \qquad 0< f < 1, \qquad A>0\end{equation}   
can arise from
\begin{equation}  V(\phi) \propto \left(\frac{\phi}{M_{\rm Pl}}\right)^{-\beta} \left( 1 - \frac{\beta^2}{6} \, \frac{M_{\rm Pl}^2}{\phi^2}\right)\end{equation} 
Thus generating a rapid expansion in the early universe is trivial if we are willing to postulate scalar fields with tailor made potentials.
This is often done in the literature.

The second point to note regarding any inflationary scenarios is that the modes with reasonable size today originated from sub-Planck length scales early on. A scale $\lambda_0$ today will be 
\begin{equation}
\lambda_{end}=\lambda_0 \frac{a_{end}}{a_0}=\lambda_0\frac{T_0}{T_{end}}\approx 
\lambda_0\times 10^{-28}
\end{equation}
at the end of inflation  (if inflation took place at GUT scales) and
\begin{equation}
\lambda_{begin}=\lambda_{end} A^{-1}\approx 
\lambda_0\times 10^{-58} (A/10^{30})^{-1}
\end{equation}
at the beginning of inflation if the inflation changed the scale factor by $A \simeq 10^{30}$. 
Note that $\lambda_{begin}<L_P$ for $\lambda_0<3$ Mpc!! Most structures in the universe today
correspond to transplanckian scales at the start of the inflation.  It is not clear whether we can
 trust standard physics at early stages of inflation or whether transplanckian effects will lead 
 to observable effects \cite{transplanck,dispersion}.

Let us get back to conventional wisdom and  consider the evolution of perturbations in a universe which underwent exponential inflation. During the inflationary phase the $a(t)$ grows exponentially and hence the wavelength of any perturbation will also grow with it. The Hubble radius, on the other hand, will remain constant. It follows that, one can have  situation in which a given mode has wavelength smaller than the Hubble radius at the beginning of the inflation but grows and becomes bigger than the Hubble radius as inflation proceeds. It is conventional to say that a perturbation of comoving wavelength $\lambda_0$  ``leaves the Hubble radius" when $\lambda_0 a=d_H$ at some time $t=t_{exit}(\lambda_0)$. For $t>t_{exit}$ the wavelength of the  perturbation is bigger than the Hubble radius. Eventually the inflation ends and the universe becomes radiation dominated. Then the wavelength will grow ($\propto t^{1/2})$ slower than the Hubble radius
($\propto t$) and will enter the Hubble radius again during $t=t_{enter}(\lambda_0)$. Our first task is to 
relate the amplitude of the  perturbation at $t=t_{exit}(\lambda_0)$ with the perturbation at $t=t_{enter}(\lambda_0)$.

We know that for modes bigger than Hubble radius, we have the conserved quantity (see \eq{zetaeqn}
\begin{equation}
   \zeta = \frac{2}{3} \frac{\rho}{\rho+p} \frac{a}{\dot a} \left( \dot\Phi + \frac{\dot a}{a}\Phi\right) +\Phi =\frac{H}{\rho+p}\frac{ik^\alpha}{k^2}\delta T^0_\alpha +\Phi
   \end{equation}
   At the time of re-entry, the universe is radiation dominated and $\zeta_{\rm entry} \approx (2/3) \Phi$.
    On the other hand, during inflation, we can write the scalar field as a dominant homogeneous part plus a small, spatially varying fluctuation: $\phi(t, {\bf x}) = \phi_0(t) + f(t, {\bf x})$.
    Perturbing the equation in \eq{onefivefive} for the scalar field, we find that 
the homogeneous mode $\phi_0$ satisfies \eq{onefivefive} while the perturbation,
in Fourier space satisfies:    
    \begin{equation}
  \frac{d^2 f_k}{dt^2} + 3 H \frac{df_k}{dt} + \frac{k^2}{a^2} f_k =0
  \end{equation}
  Further, the energy momentum tensor for the scalar field gives
 [with  the  ``dot'' denoting $(d/d\eta) = a(d/dt)$]:
 \begin{equation}
  \rho = \frac{\dot \phi_0^2}{2 a^2} + V; \quad \ p = \frac{\dot \phi_0^2}{2 a^2} - V
  ; \quad \delta T^\alpha_0 = \frac{ik^\alpha}{a} \dot \phi_0 \, f
  \end{equation}
   It is easy to see that
 $\Phi $ is negligible
   at $t=t_{\rm exit}$ since  
   \begin{equation}
    \Phi \sim \frac{4\pi G a^2}{k^2} \sim \frac{4\pi G }{H^2}\delta \rho \sim \frac{\delta \rho}{\rho} \sim \left(\frac{\rho+p}{\rho}\right)\left( \frac{\delta \rho}{\rho+p}\right) \ll \frac{H}{(\rho+p)} \frac{\dot \phi_0}{a} \, f_k
   \end{equation}
 Therefore, 
   \begin{equation}
   \zeta_{\rm exit} \approx  \frac{H}{(\dot \phi_0^2/a^2)} \left[ - \frac{\dot \phi_0 f_k}{a}\right] = -aH\frac{f_k}{\dot \phi_0} \approx \frac{3H^2}{V'} f_k
   \end{equation}
Using the conservation law $\zeta_{exit}=\zeta_{entry},$ we get
\begin{equation}
\Phi_k\Big|_{\rm entry}=\frac{9H^2}{2V'}f_k\Big|_{\rm exit}
\label{phikentry}
\end{equation}   
Thus, given a perturbation of the scalar field $f_k$ during inflation, we can compute its value at the time of re-entry, which --- in turn --- can be used to compare with observations. 
 
Equation (\ref{phikentry}) connects a \textit{classical} energy density perturbation $f_k$ at the time of exit with the corresponding quantity $\Phi_k$ at the time of re-entry. The next important --- and conceptually difficult --- question is how we can obtain a  
 \textit{c-number} field $f_k$ from a quantum scalar field. There is no simple answer to this question and one possible way of doing it is as follows: Let us start with the quantum operator for a scalar field decomposed into the Fourier modes with $\hat q_{\bf k}(t)$ denoting an infinite set of operators:
\begin{equation} \hat \phi(t,{\bf x})=
\int{d^3{\bf k}\over (2\pi)^3}\hat q_{\bf k}(t){\rm e}^{i{\bf k.x}}.\end{equation}  
We choose a quantum state $|\psi>$ such the expectation value of $\hat q_{\bf k}(t)$ vanishes for all non-zero $\mathbf{k}$
 so that the expectation value of $\hat \phi(t,{\bf x})$ gives the homogeneous mode that drives the inflation. The quantum fluctuation around this homogeneous part in a quantum state $|\psi>$ is given by 
\begin{equation} \sigma^2_{\bf k}(t)=
<\psi| \hat q^2_{\bf k}(t)|\psi>-<\psi| \hat q_{\bf k}(t)|\psi>^2=
<\psi| \hat q^2_{\bf k}(t)|\psi>\end{equation}  
It is easy to verify that this fluctuation is just the Fourier transform of the two-point function in this state: 
\begin{equation} \sigma^2_{\bf k}(t)=\int d^3{\bf x}<\psi|\hat \phi(t,{\bf x+y})
\hat\phi(t,{\bf y})|\psi>{\rm e}^{i{\bf k.x}}.\end{equation}  
Since $\sigma_{\bf k}$ characterises the quantum fluctuations, it seems reasonable to introduce a c-number field $f(t,{\bf x})$ by the definition:
\begin{equation}  f(t,{\bf x})\equiv\int{d^3 {\bf k}\over (2\pi)^3}
\sigma_{\bf k}(t){\rm e}^{i{\bf k.x}}
\label{cnumber}
\end{equation} 
This c-number field will have same \textit{c-number power spectrum} as the \textit{quantum} fluctuations. Hence we may take this as our definition of an equivalent classical perturbation. (There are more sophisticated ways of getting this result but none of them are fundamentally more sound that the elementary definition given above. There is a large literature on the subject of quantum to classical transition, especially in the context of gravity; see e.g.\cite{semicgrav})
We now have all the ingredients in place. Given the quantum state $|\psi>$, one can explicitly compute $\sigma_{\bf k}$ and then --- using \eq{phikentry} with $f_\mathbf{k} = \sigma_\mathbf{k}$ --- obtain the density perturbations at the time of re-entry.

The next question we need to address is what is $|\psi>$. The free quantum field theory in the 
 Friedmann background  is identical to the  quantum mechanics of a bunch of time dependent harmonic oscillators, each labelled by a wave vector ${\bf k}$. 
The action for a free scalar field in the Friedmann background 
\begin{equation}
A=\frac{1}{2} \int d^4x\, \sqrt{-g}\, \partial_a\phi \partial^a\phi
= \frac{1}{2} \int dt\, d^3{\bf x}\,  a^3 \left[ \left(\frac{\partial\phi}{\partial t}\right)^2 - \frac{1}{a^2}(\nabla\phi)^2
\right] 
\rightarrow  \frac{1}{2} \int dt\, d^3{\bf k}\,  a^3 \left[ \left(\frac{dq_{\bf k}}{dt}\right)^2 - \frac{k^2}{a^2}q_{\bf k}^2 \right]
\end{equation} 
can be thought of as the sum over the actions for an infinite set of harmonic oscillators with
mass $m= a^3$ and frequency $\omega_k^2 = k^2/a^2$. (To be precise, one needs to treat the real and imaginary parts of the Fourier transform as independent oscillators and restrict the range of ${\bf k}$; just pretending that $q_{\bf k}$ is real amounts the same thing.) The quantum state of the field is just an 
infinite product of the quantum state $\psi_{\bf k}[q_{\bf k}, t]$ for each of the harmonic
oscillators and satisfies the Schrodinger equation
\begin{equation}
i\frac{\partial\psi_{\bf k}}{\partial t} = - \frac{1}{2a^3}\frac{\partial^2 \psi_{\bf k}}{\partial q_{\bf k}^2} + \frac{1}{2} ak^2\psi_{\bf k}
\label{se}
\end{equation} 
If the quantum state $\psi_{\bf k}[q_{\bf k}, t']$ of any given oscillator, labelled by ${\bf k}$, is given at some initial time, $t'$,  we can evolve it to final time:
 \begin{equation} 
\psi \left[ q_{\bf k}, t\right] = \int dq_{\bf k}' \, K[q_{\bf k}, t; q_{\bf k}', t'] \psi \left[ q_{\bf k}', t'\right]
\end{equation}
where \( K\) is known in terms of the solutions to the classical equations of motion and
\( \psi [ q_{\bf k}', t'] \) is the initial state. 
There is nothing non-trivial in the mathematics, but the physics is completely unknown.
The real problem is that 
unfortunately  --- in spite of confident assertions in the literature occasionally --- we have no clue  what \( \psi [ q_{\bf k}', t'] \) is. 
So we need to make more assumptions to proceed further. 

One natural choice is the following: It  turns out that, 
 Gaussian states  of the form
  \begin{equation}
  \psi_{\bf k} = A_{\bf k} (t) \exp[ - B_{\bf k} (t) q_{\bf k}^2]
  \label{gs}
  \end{equation}
  preserve their form under evolution governed by the Schrodinger equation in \eq{se}.
  Substituting \eq{gs} in \eq{se} we can determine the ordinary differential equation
  which governs $B_{\bf k} (t)$. (The $A_{\bf k} (t)$ is trivially fixed by normalization.)
  Simple algebra shows that $ B_{\bf k}(t)\) can be expressed in the form 
 \begin{equation} 
   B_{\bf k} = - \frac{i}{2} a^3 \left( \frac{1}{f_{\bf k}}\frac{df_{\bf k}}{dt}\right)
  \end{equation} 
 where $f_k$ is the  solution to the classical equation of motion: 
   \begin{equation}
  \frac{d^2 f_k}{dt^2} + 3 H \frac{df_k}{dt} + \frac{k^2}{a^2} f_k =0
  \label{onesevenseven}
  \end{equation}
For the quantum state in \eq{gs}, the fluctuations are characterized by
\begin{equation} 
   \sigma_{\bf k}^2 = \frac{1}{2} ( {\rm Re}\, B_{\bf k} )^{-1} = |f_{\bf k}|^2
  \end{equation}
   
 Since one can take different choices for the solutions of \eq{onesevenseven} one get different values for $\sigma_{\bf k}$ and different spectra for perturbations. Any prediction one makes depends on the choice of mode functions. One possibility is to choose the modes so that  $\psi_{\bf k}$ represents  the instantaneous vacuum state of the oscillators at some time $t=t_i$. 
 (That is Re $B_{\bf k}(t_i)=(1/2)\omega_{\bf k}^2(t_i),$ say). The final result will then depend on the choice for $t_i$. One can further make an assumption that we are interested in the limit of $t_i\to -\infty$; that is the quantum state is an instantaneous ground state in the infinite past. It is easy to show that this corresponds to choosing the following solution to \eq{onesevenseven}:
   \begin{equation} 
  f_k =\frac{1}{a\sqrt{2k}}(1+ ix) e^{ix}; \qquad x = \frac{k}{Ha} 
  \end{equation}
 which is usually called the Bunch-Davies vacuum. 
 For this choice,
\begin{equation}
|f_k|^2=\frac{1}{2ka^2}\left(1+\frac{k^2}{a^2H^2}\right); \quad|f_k|^2\big|_{k=aH}\approx \frac{H^2}{k^3}
\end{equation}
where the second result is at $t = t_{\rm exit}$ which is what we need to use in \eq{phikentry}, (Numerical factors of order unity cannot be trusted in this computation). We can now determine the amplitude of the perturbation when it re-enters the Hubble radius. \eq{phikentry} gives:
\begin{equation}
|\Phi_k|^2_{entry}=\left(\frac{9H^2}{2V'}\right)^2|f_k|^2=\frac{1}{k^3}\frac{9H^6}{4V'^2}; \qquad k^3 |\Phi_k|^2_{\rm entry} \simeq \left( \frac{H^3}{V'}\right)^2_{\rm exit}
\label{oneeightone}
\end{equation}
 One sees that the result is scale invariant in the sense that
 $k^3|\Phi_k|_{\rm entry}^2$ is independent of $k$.
 
 It is sometimes claimed in the literature that scale invariant spectrum is a prediction of inflation. \textit{This is simply wrong.} One has to make several \textit{other} assumptions including an all important choice for the quantum state (about which we know nothing) to obtain scale invariant spectrum. In fact, one can prove that, given any power spectrum $\Phi(k)$, one can find a quantum state such that this power spectrum is generated (for an explicit construction, see the last reference in \cite{transplanck}). So whatever results are obtained by observations can be reconciled with inflationary generation of perturbations. 
 
To conclude the discussion, let us work out the perturbations for one specific case. 
Let us consider the case of the $\lambda \phi^4$ model for which
  \begin{equation} 
  \frac{d^2 \phi}{dt^2}  + 3 H \frac{d\phi}{dt} + V'(\phi) =0; \quad V(\phi) \approx V_0 - {\lambda\over 4}\phi^4
  \end{equation}
Using
\begin{equation} N  \cong 8\pi G \int_\phi^{\phi_f} {V_0\over [-V']} d\phi = {3H^2 \over 2\lambda} \left( {1\over \phi^2} - {1\over \phi^2_f}\right) \approx {3H^2 \over 2\lambda \phi^2}\end{equation}   
we can write
\begin{equation} -V'(\phi) = \lambda \phi^3 \simeq {H^3\over \lambda^{1/2} N^{3/2}}.\end{equation}   
so that the result in \eq{oneeightone} becomes:
\begin{equation} k^{3/2} \Phi_k \simeq H^3  {\lambda^{1/2} N^{3/2} \over H^3} \approx  \lambda^{1/2} N^{3/2}.\end{equation}  
We do get scale invariant spectrum but the amplitude has a serious problem. If we take
$ N\gtrsim 50$ and note that observations require $k^{3/2} \Phi_k \sim 10^{-4}$ we need to take \( \lambda \lesssim 10^{-15}\)
for getting consistent values. Such a fine tuning of a dimensionless coupling constant is fairly ridiculous; but over years inflationists have learnt to successfully forget this embarrassment.

Our formalism can also be used to estimate the deviation of the power spectrum from the scale invariant form. To the lowest order we
have 
\begin{equation}  
\Delta_{\Phi}^2 \sim k^3|\Phi_k|^2 \sim \frac{H^6}{(V')^2} \sim \left(\frac{V^3}{m_P^6V^{'2}}\right)\end{equation}  
Let us define the deviation from the scale invariant index by
 \( (n-1) = (d \ln \Delta_\phi^2 /d \ln k)\). Using
\begin{equation}  \frac{d}{d \ln k} =a \frac{d}{da} = \frac{\dot \phi}{H} \frac{d}{d\phi} = - \frac{m_p^2}{8\pi} \frac{V'}{V} \frac{d}{d\phi}\end{equation}  
one finds that
\begin{equation} 1-n = 6\epsilon - 2\eta
\label{oneeightnine}
\end{equation}
Thus, as long as $\epsilon$  and $\eta$ are small we do have $n\approx 1$; what is more, given a potential one can estimate 
$\epsilon$  and $\eta$ and thus the deviation $(n-1)$.

Finally, note that the same  process can also generate spin-2 perturbations.  If we take the normalised gravity wave amplitude as \( h_{ab} = \sqrt{16 \pi G}\, e_{ab}\phi\), the mode function
   \(\phi\) behaves like a scalar field.
    (The normalisation is dictated by the fact that the action for the perturbation should reduce to that of a spin-2 field.)
    The corresponding power spectrum of gravity waves is
   \begin{equation} P_{\rm grav}(k)\cong
{k^3|h_k|^2\over 2\pi^2}=
{4\over \pi}
\left({H\over m_P}\right)^2, 
\quad
\Omega_{\rm grav}(k)h^2\simeq 10^{-5}
\left({M\over m_P}\right)^3 \end{equation}  
Comparing the two results
\begin{equation}  \Delta_{\rm scalar}^2  \sim \frac{H^6}{(V')^2} \sim \left(\frac{V^3}{m_P^6V^{'2}}\right); \quad  \Delta_{\rm tensor}^2  \sim \left(\frac{H^2}{m_P^2}\right)\sim \left(\frac{V}{m_P^4}\right)\end{equation} 
we get \( (\Delta _{\rm tensor}/\Delta_{\rm scalar})^2 \approx 16 \pi \epsilon \ll 1\). Further, if \( (1-n) \approx 4\epsilon\)  (see \eq{oneeightnine} with $
\epsilon \sim \eta$) we have
the relation
\( (\Delta_{\rm tensor}/\Delta_{\rm scalar})^2 \approx \mathcal{O}(3) (1-n)\) which  connects three quantities, all of which are independently observable in principle. If these are actually measured in future it could act as a consistency check of the inflationary paradigm.

\section{The Dark Energy}

It is rather frustrating that the only component of the universe which we understand theoretically is the radiation! While understanding the
baryonic and dark matter components [in particular the values of $\Omega_B$ and $\Omega_{DM}$] is by no means trivial, the issue of dark energy is lot more perplexing, thereby justifying the attention it has received recently. In this section we will discuss several aspects of the dark energy problem.

The key observational feature of dark energy is that --- treated as a fluid with a stress tensor $T^a_b=$ dia     $(\rho, -p, -p,-p)$ 
--- it has an equation state $p=w\rho$ with $w \lesssim -0.8$ at the present epoch. 
The spatial part  ${\bf g}$  of the geodesic acceleration (which measures the 
  relative acceleration of two geodesics in the spacetime) satisfies an \textit{exact} equation
  in general relativity  given by:
  \begin{equation}
  \nabla \cdot {\bf g} = - 4\pi G (\rho + 3p)
  \label{nextnine}
  \end{equation} 
 This  shows that the source of geodesic  acceleration is $(\rho + 3p)$ and not $\rho$.
  As long as $(\rho + 3p) > 0$, gravity remains attractive while $(\rho + 3p) <0$ can
  lead to repulsive gravitational effects. In other words, dark energy with sufficiently negative pressure will
  accelerate the expansion of the universe, once it starts dominating over the normal matter.  This is precisely what is established from the study of high redshift supernova, which can be used to determine the expansion
rate of the universe in the past \cite{sn}. 

The simplest model for  a fluid with negative pressure is the
cosmological constant (for some recent  reviews, see \cite{cc}) with $w=-1,\rho =-p=$ constant.
If the dark energy is indeed a cosmological constant, then it introduces a fundamental length scale in the theory $L_\Lambda\equiv H_\Lambda^{-1}$, related to the constant dark energy density $\rho_{_{\rm DE}}$ by 
$H_\Lambda^2\equiv (8\pi G\rho_{_{\rm DE}}/3)$.
In classical general relativity,
    based on the constants $G, c $ and $L_\Lambda$,  it
  is not possible to construct any dimensionless combination from these constants. But when one introduces the Planck constant, $\hbar$, it is  possible
  to form the dimensionless combination $H^2_\Lambda(G\hbar/c^3) \equiv  (L_P^2/L_\Lambda^2)$.
  Observations then require $(L_P^2/L_\Lambda^2) \lesssim 10^{-123}$.
  As has been mentioned several times in literature, this will require enormous fine tuning. What is more,
 in the past, the energy density of 
  normal matter and radiation  would have been higher while the energy density contributed by the  cosmological constant
  does not change.  Hence we need to adjust the energy densities
  of normal matter and cosmological constant in the early epoch very carefully so that
  $\rho_\Lambda\gtrsim \rho_{\rm NR}$ around the current epoch.
  This raises the second of the two cosmological constant problems:
  Why is it that $(\rho_\Lambda/ \rho_{\rm NR}) = \mathcal{O} (1)$ at the 
  {\it current} phase of the universe ?

  Because of these conceptual problems associated with the cosmological constant, people have explored a large variety of alternative possibilities. The most popular among them uses a scalar field $\phi$ with a suitably chosen potential $V(\phi)$ so as to make the vacuum energy vary with time. The hope then is that, one can find a model in which the current value can be explained naturally without any fine tuning.
  A simple form of the source with variable $w$ are   scalar fields with
  Lagrangians of different forms, of which we will discuss two possibilities:
    \begin{equation}
  L_{\rm quin} = \frac{1}{2} \partial_a \phi \partial^a \phi - V(\phi); \quad L_{\rm tach}
  = -V(\phi) [1-\partial_a\phi\partial^a\phi]^{1/2}
  \label{lquineq}
  \end{equation}
  Both these Lagrangians involve one arbitrary function $V(\phi)$. The first one,
  $L_{\rm quin}$,  which is a natural generalization of the Lagrangian for
  a non-relativistic particle, $L=(1/2)\dot q^2 -V(q)$, is usually called quintessence (for
  a small sample of models, see \cite{phiindustry}; there is an extensive and growing literature on scalar field models and more references can be found in the reviews in ref.\cite{cc}).
    When it acts as a source in Friedmann universe,
   it is characterized by a time dependent
  $w(t)$ with
    \begin{equation}
  \rho_q(t) = \frac{1}{2} \dot\phi^2 + V; \quad p_q(t) = \frac{1}{2} \dot\phi^2 - V; \quad w_q
  = \frac{1-(2V/\dot\phi^2)}{1+ (2V/\dot\phi^2)}
  \label{quintdetail}
  \end{equation}

The structure of the second Lagrangian  (which arise in string theory \cite{asen}) in Eq.~(\ref{lquineq}) can be understood by a simple analogy from
special relativity. A relativistic particle with  (one dimensional) position
$q(t)$ and mass $m$ is described by the Lagrangian $L = -m \sqrt{1-\dot q^2}$.
It has the energy $E = m/  \sqrt{1-\dot q^2}$ and momentum $k = m \dot
q/\sqrt{1-\dot q^2} $ which are related by $E^2 = k^2 + m^2$.  As is well
known, this allows the possibility of having \textit{massless} particles with finite
energy for which $E^2=k^2$. This is achieved by taking the limit of $m \to 0$
and $\dot q \to 1$, while keeping the ratio in $E = m/  \sqrt{1-\dot q^2}$
finite.  The momentum acquires a life of its own,  unconnected with the
velocity  $\dot q$, and the energy is expressed in terms of the  momentum
(rather than in terms of $\dot q$)  in the Hamiltonian formulation. We can now
construct a field theory by upgrading $q(t)$ to a field $\phi$. Relativistic
invariance now  requires $\phi $ to depend on both space and time [$\phi =
\phi(t, {\bf x})$] and $\dot q^2$ to be replaced by $\partial_i \phi \partial^i
\phi$. It is also possible now to treat the mass parameter $m$ as a function of
$\phi$, say, $V(\phi)$ thereby obtaining a field theoretic Lagrangian $L =-
V(\phi) \sqrt{1 - \del^i \phi \del_i \phi}$. The Hamiltonian  structure of this
theory is algebraically very similar to the special  relativistic example  we
started with. In particular, the theory allows solutions in which $V\to 0$,
$\dphi \to 1$ simultaneously, keeping the energy (density) finite.  Such
solutions will have finite momentum density (analogous to a massless particle
with finite  momentum $k$) and energy density. Since the solutions can now
depend on both space and time (unlike the special relativistic example in which
$q$ depended only on time), the momentum density can be an arbitrary function
of the spatial coordinate. The structure of this Lagrangian is similar to those analyzed in a wide class of models
   called {\it K-essence} \cite{kessence} and  provides a rich gamut of possibilities in the
context of cosmology
 \cite{tptachyon,tachyon}.

   Since  the quintessence field (or the tachyonic field)   has
   an undetermined free function $V(\phi)$, it is possible to choose this function
  in order to produce a given $H(a)$.
  To see this explicitly, let
   us assume that the universe has two forms of energy density with $\rho(a) =\rho_{\rm known}
  (a) + \rho_\phi(a)$ where $\rho_{\rm known}(a)$ arises from any known forms of source 
  (matter, radiation, ...) and
  $\rho_\phi(a) $ is due to a scalar field.  
  Let us first consider quintessence. Here,  the potential is given implicitly by the form
  \cite{ellis,tptachyon}. 
  \begin{equation}
  V(a) = \frac{1}{16\pi G} H (1-Q)\left[6H + 2aH' - \frac{aH Q'}{1-Q}\right];
  \quad
    \phi (a) =  \left[ \frac{1}{8\pi G}\right]^{1/2} \int \frac{da}{a}
     \left[ aQ' - (1-Q)\frac{d \ln H^2}{d\ln a}\right]^{1/2}
    \label{phioft}
    \end{equation} 
   where $Q (a) \equiv [8\pi G \rho_{\rm known}(a) / 3H^2(a)]$ and prime denotes differentiation with respect to $a$.
   (The result used in \eq{vteqn} is just a special case of this when $Q=0$)
   Given any
   $H(a),Q(a)$, these equations determine $V(a)$ and $\phi(a)$ and thus the potential $V(\phi)$. 
   \textit{Every quintessence model studied in the literature can be obtained from these equations.}
  
  Similar results exists for the tachyonic scalar field as well \cite{tptachyon}. For example, given
  any $H(a)$, one can construct a tachyonic potential $V(\phi)$ so that the scalar field is the 
  source for the cosmology. The equations determining $V(\phi)$  are now given by:
  \begin{equation}
  \phi(a) = \int \frac{da}{aH} \left(\frac{aQ'}{3(1-Q)}
   -{2\over 3}{a H'\over H}\right)^{1/2};
  \quad
   V = {3H^2 \over 8\pi G}(1-Q) \left( 1 + {2\over 3}{a H'\over H}-\frac{aQ'}{3(1-Q)}\right)^{1/2}
   \label{finalone}
   \end{equation}
   Equations (\ref{finalone}) completely solve the problem. Given any
   $H(a)$, these equations determine $V(a)$ and $\phi(a)$ and thus the potential $V(\phi)$. 
A wide variety of phenomenological models with time dependent
  \cc\ have been considered in the literature all of which can be 
   mapped to a 
  scalar field model with a suitable $V(\phi)$.

  While the scalar field models enjoy considerable popularity (one reason being they are easy to construct!)
  it is very doubtful whether they have helped us to understand the nature of the dark energy
  at any deeper level. These
  models, viewed objectively, suffer from several shortcomings:
  \begin{figure}[ht]
 \begin{center}
 \includegraphics[scale=0.6]{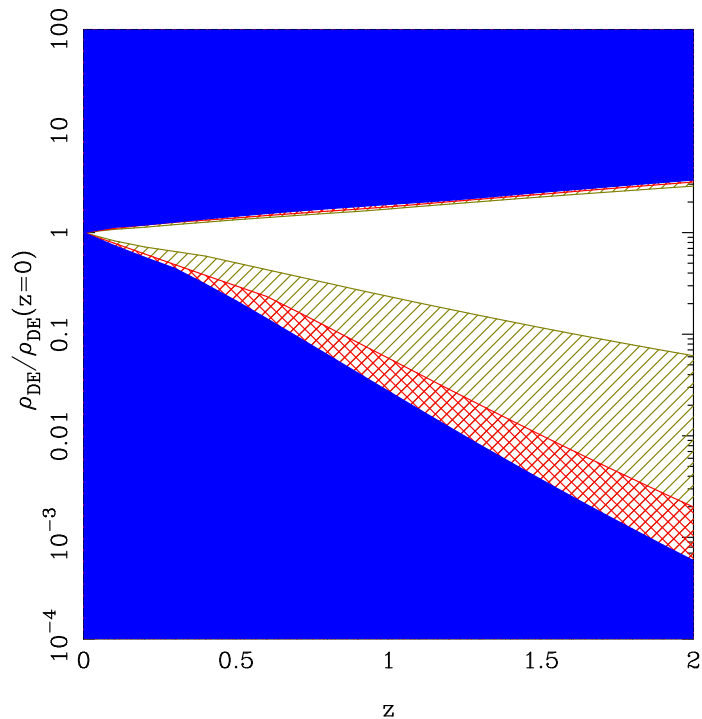}
 \end{center}
 \caption{
 The observational constraints on the variation of dark energy density  as a function of
 redshift  from WMAP and SNLS data (see \cite{jbp}). The green/hatched region is
 excluded  at $68\%$ confidence limit, red/cross-hatched region at $95\%$
 confidence level  and the blue/solid region at $99\%$ confidence limit. The
 white region shows the allowed range of variation of dark energy at $68\%$
 confidence limit.  
  }
 \label{fig:bjp2ps}
 \end{figure}  
  \begin{itemize}
  \item
  They completely lack predictive power. As explicitly demonstrated above, virtually every form of $a(t)$ can be modeled by a suitable ``designer" $V(\phi)$.
  \item
  These models are  degenerate in another sense. The previous discussion  illustrates that even when $w(a)$ is known/specified, it is not possible to proceed further and determine
  the nature of the scalar field Lagrangian. The explicit examples given above show that there
  are {\em at least} two different forms of scalar field Lagrangians (corresponding to
  the quintessence or the tachyonic field) which could lead to
  the same $w(a)$. (See the second paper in ref.\cite{tptirthsn1} for an explicit example of such a construction.)
  \item
  All the scalar field potentials require fine tuning of the parameters in order to be viable. This is obvious in the quintessence models in which adding a constant to the potential is the same as invoking a \cc. So to make the quintessence models work, \textit{we first need to assume the \cc\ is zero.} These models, therefore, merely push the cosmological constant problem to another level, making it somebody else's problem!.
  \item
  By and large, the potentials  used in the literature have no natural field theoretical justification. All of them are non-renormalisable in the conventional sense and have to be interpreted as a low energy effective potential in an ad hoc manner.
  \item
  One key difference between \cc\ and scalar field models is that the latter lead to a $w(a)$ which varies with time. If observations have demanded this, or even if observations have ruled out $w=-1$ at the present epoch,
  then one would have been forced to take alternative models seriously. However, all available observations are consistent with \cc\ ($w=-1$) and --- in fact --- the possible variation of $w$ is strongly constrained \cite{jbp} as shown in Figure \ref{fig:bjp2ps}. 
  \item
  While on the topic of observational constraints on $w(t)$, it must be stressed that: (a) There is fair amount of tension between WMAP and SN data  and one should be very careful about the priors used in these analysis. (b) There is no observational evidence for $w<-1$. (c) It is likely that more homogeneous, future, data sets of SN might show better agreement with WMAP results. (For more details related to these issues, see the last reference in \cite{jbp}.)
 \end{itemize}

The observational and theoretical features described above suggests that one should consider \cc\ as the most natural candidate for dark energy. Though it leads to well know fine tuning problems, it also has certain attractive features that need to kept in mind.
\begin{itemize}
\item
Cosmological constant is the most economical [just one number] and simplest  explanation for all the observations. We stress  that there is absolutely \textit{no} evidence for variation of dark energy density with redshift, which is consistent with the assumption of \cc\ .
\item
Once we invoke the \cc\ classical gravity will be described by the three constants $G,c$ and $\Lambda\equiv L_\Lambda^{-2}$. It is \textit{not} possible to obtain a dimensionless quantity from these; so, within classical theory, there is no fine tuning issue. Since $\Lambda(G\hbar/c^3)\equiv (L_P/L_\Lambda)^2\approx 10^{-123}$, it is obvious that the \cc\ is telling us something regarding \textit{quantum gravity}, indicated by the combination $G\hbar$. \textit{An acid test for any quantum gravity model will be its ability to explain this value;} needless to say, all the currently available models --- strings, loops etc.  --- flunk this test.
\item
So, if dark energy is indeed \cc\, this will be the greatest contribution from cosmology to fundamental physics. It will be unfortunate if we miss this chance by invoking some scalar field epicycles!
\end{itemize}

In this context, it is worth  stressing another   peculiar feature of \cc\, when it is  treated as a clue to quantum gravity.
It is well known that, based on energy scales, the \cc\ problem is an infra red problem \textit{par excellence}.
At the same time, it is a relic of a quantum gravitational effect or principle of unknown nature. An analogy \cite{choices} will be helpful to illustrate this point. Suppose one solves the Schrodinger equation for the Helium atom for the quantum states of the two electrons $\psi(x_1,x_2)$. When the result is compared with observations, one will find that only half the states --- those in which  $\psi(x_1,x_2)$ is antisymmetric under $x_1\longleftrightarrow x_2$ interchange --- are realized in nature. But the low energy Hamiltonian for electrons in the Helium atom has no information about
this effect! Here is low energy (IR) effect which is a relic of relativistic quantum field theory (spin-statistics theorem) that is  totally non perturbative, in the sense that writing corrections to the Helium atom Hamiltonian in some $(1/c)$ expansion will {\it not} reproduce this result. The current value of \cc\ could very well be related to quantum gravity in a similar way. There must exist a deep principle in quantum gravity which leaves its non perturbative trace even in the low energy limit
that appears as the \cc\ .

Let us now turn our attention to few of the many attempts to understand the \cc. The choice  is, of course, dictated by personal bias and is definitely a non-representative sample. A host of other approaches exist in literature, some of which can be found in \cite{catchall}.

\subsection{Gravitational Holography}

One possible way of addressing this issue is to simply eliminate from the gravitational theory those modes which couple to cosmological constant. If, for example, we have a theory in which the source of gravity is
$(\rho +p)$ rather than $(\rho +3p)$ in Eq. (\ref{nextnine}), then \cc\ will not couple to gravity at all. (The non linear coupling of matter with gravity has several subtleties; see eg. \cite{gravitonmyth}.) Unfortunately
it is not possible to develop a covariant theory of gravity using $(\rho +p)$ as the source. But we can probably gain some insight from the following considerations. Any metric $g_{ab}$ can be expressed in the form $g_{ab}=f^2(x)q_{ab}$ such that
${\rm det}\, q=1$ so that ${\rm det}\, g=f^4$. From the action functional for gravity
\begin{equation}
A=\frac{1}{16\pi G}\int d^4x (R -2\Lambda)\sqrt{-g}
=\frac{1}{16\pi G}\int d^4x R \sqrt{-g}-\frac{\Lambda}{8\pi G}\int d^4x f^4(x)
\end{equation}
it is obvious that the \cc\ couples {\it only} to the conformal factor $f$. So if we consider a theory of gravity in which $f^4=\sqrt{-g}$ is kept constant and only $q_{ab}$ is varied, then such a model will be oblivious of
direct coupling to \cc. If the action (without the $\Lambda$ term) is varied, keeping ${\rm det}\, g=-1$, say, then one is lead to a {\it unimodular theory of gravity} that has  the equations of motion 
$R_{ab}-(1/4)g_{ab}R=\kappa(T_{ab}-(1/4)g_{ab}T)$ with zero trace on both sides. Using the Bianchi identity, it is now easy to show that this is equivalent to the usual  theory with an {\it  arbitrary} \cc. That is, \cc\ arises as an undetermined integration constant in this model \cite{unimod}. 

The same result arises in another, completely different approach to gravity. In the standard approach to gravity one uses the Einstein-Hilbert Lagrangian  $L_{EH}\propto R$ which has a formal structure $L_{EH}\sim R\sim (\partial g)^2+\partial^2g$. 
If the surface term obtained by integrating $L_{sur}\propto \partial^2g$ is ignored (or, more formally, canceled by an extrinsic curvature term) then the Einstein's equations arise from the variation of the bulk
term $L_{bulk}\propto (\partial g)^2$ which is the non-covariant $\Gamma^2$ Lagrangian.
There is, however, a remarkable relation 
between $L_{bulk}$ and $L_{sur}$:
\begin{equation}
\sqrt{-g}L_{sur}=-\partial_a\left(g_{ij}
\frac{\partial \sqrt{-g}L_{bulk}}{\partial(\partial_ag_{ij})}\right)
\end{equation}
which allows a dual description of gravity using either $L_{bulk}$ or $L_{sur}$!
It is possible to obtain  the dynamics of gravity \cite{tpholo} from an approach which  uses \textit{only} the surface term of the Hilbert action; \textit{ we do not need the bulk term at all!}. This suggests that \textit{the true  degrees of freedom of gravity
for a volume $\mathcal{V}$
reside in its boundary $\partial\mathcal{V}$} --- a  point of view that is strongly supported by the study
of horizon entropy, which shows that the degrees of freedom hidden by a horizon scales as the area and not as the volume. 
The resulting equations can be cast
in a thermodynamic form $TdS=dE+PdV$ and the
 continuum spacetime is like an elastic solid (see e.g. \cite{sakharov}) with Einstein's equations providing the macroscopic description. Interestingly, the \textit{\cc\ arises again in this approach as a undetermined integration constant} but closely related to the `bulk expansion' of the solid.

While this is all very interesting, we still need an extra physical principle to fix the value (even the sign) of \cc.
One possible way of doing this is to  interpret the $\Lambda$ term in the action as a Lagrange multiplier for the proper volume of the spacetime. Then it is reasonable to choose the \cc\ such that the total proper volume of the universe is equal to a specified number. While this will lead to a \cc\ which has the correct order of magnitude, it has several obvious problems. First, the proper four volume of the universe is infinite unless we make the spatial sections compact and restrict the range of time integration. Second, this will lead to a dark energy density  which varies as $t^{-2}$ (corresponding to $w= -1/3$ ) which is ruled out by observations. 

\subsection{Cosmic Lenz law}

Another possibility which has been attempted in the literature tries to ``cancel out'' the \cc\ by some process,
usually quantum mechanical in origin. One of the simplest ideas will be to ask whether switching on a \cc\ will
lead to a vacuum polarization with an effective energy momentum tensor that will tend to cancel out the \cc.
A less subtle way of doing this is to invoke another scalar field (here we go again!) such that it can couple to 
\cc\ and reduce its effective value \cite{lenz}. Unfortunately, none of this could be made to work properly. By and large, these approaches lead to an energy density which is either $\rho_{_{\rm UV}}\propto L_P^{-4}$ (where 
 $L_P$ is the Planck length) or to $\rho_{_{\rm IR}}\propto L_\Lambda^{-4}$ (where 
 $L_\Lambda=H_\Lambda^{-1}$ is the Hubble radius associated with the \cc\ ). The first one is too large while the second one is too small! 
 
 \subsection{Geometrical Duality in our Universe}
 
 While the above ideas do not work, it gives us a clue. A universe with two
 length scales $L_\Lambda$ and $L_P$  will be  asymptotically De Sitter with $a(t)\propto \exp (t/L_\Lambda) $ at late times. There are some curious features in such a universe which we will now describe.  Given the two length scales $L_P$ and $L_\Lambda$, one can construct two energy scales
 $\rho_{_{\rm UV}}=1/L_P^4$ and $\rho_{_{\rm IR}}=1/L_\Lambda^4$ in natural units ($c=\hbar=1$). There is sufficient amount of justification from different theoretical perspectives
 to treat $L_P$ as the zero point length of spacetime \cite{zeropoint}, giving a natural interpretation to $\rho_{_{\rm UV}}$. The second one, $\rho_{_{\rm IR}}$ also has a natural interpretation. The universe which is asymptotically De Sitter has a horizon and associated thermodynamics \cite{ghds} with a  temperature
 $T=H_\Lambda/2\pi$ and the corresponding thermal energy density $\rho_{thermal}\propto T^4\propto 1/L_\Lambda^4=
 \rho_{_{\rm IR}}$. Thus $L_P$ determines the \textit{highest} possible energy density in the universe while $L_\Lambda$
 determines the {\it lowest} possible energy density in this universe. As the energy density of normal matter drops below this value, the thermal ambience of the De Sitter phase will remain constant and provide the irreducible `vacuum noise'. \textit{Note that the dark energy density is the the geometric mean $\rho_{_{\rm DE}}=\sqrt{\rho_{_{\rm IR}}\rho_{_{\rm UV}}}$ between the two energy densities.} If we define a dark energy length scale $L_{DE}$  such that $\rho_{_{\rm DE}}=1/L_{DE}^4$ then $L_{DE}=\sqrt{L_PL_\Lambda}$ is the geometric mean of the two length scales in the universe. (Incidentally, $L_{DE}\approx 0.04$ mm is macroscopic; it is also pretty close to the length scale associated with a neutrino mass of $10^{-2}$ eV; another intriguing coincidence ?!)

  Using the characteristic length scale of expansion,
 the Hubble radius $d_H\equiv (\dot a/a)^{-1}$, we can distinguish between three different phases of such a universe. The first phase is when the universe went through a inflationary expansion with $d_H=$ constant; the second phase is the radiation/matter dominated phase in which most of the standard cosmology operates and $d_H$ increases monotonically; the third phase is that of re-inflation (or accelerated expansion) governed by the cosmological constant in which $d_H$ is again a constant. The first and last phases are time translation invariant;
 that is, $t\to t+$ constant is an (approximate) invariance for the universe in these two phases. The universe satisfies the perfect cosmological principle and is in steady state during these phases!
 
 In fact, one can easily imagine a scenario in which the two De Sitter phases (first and last) are of arbitrarily long duration \cite{plumian}. If  $\Omega_\Lambda\approx 0.7, \Omega_{DM}\approx 0.3$ the final De Sitter phase \textit{does} last forever; as regards the inflationary phase, nothing prevents it from lasting for arbitrarily long duration. Viewed from this perspective, the in between phase --- in which most of the `interesting' cosmological phenomena occur ---  is  of negligible measure in the span of time. It merely connects two steady state phases of the universe.

 While the two De Sitter phases can last forever in principle, there is a natural cut off length scale in both of them
 which makes the region of physical relevance to be finite \cite{plumian}. Let us first discuss the case of re-inflation in the late universe. 
 As the universe grows exponentially in the phase 3, the wavelength of CMBR photons are being redshifted rapidly. When the temperature of the CMBR radiation drops below the De Sitter temperature (which happens when the wavelength of the typical CMBR photon is stretched to the $L_\Lambda$.)
 the universe will be essentially dominated by the vacuum thermal noise \cite{ghds} due to the horizon in the De Sitter phase.
 This happens  when the expansion factor is $a=a_F$ determined by the
  equation $T_0 (a_0/a_{F}) = (1/2\pi L_\Lambda)$. Let $a=a_\Lambda$ be the epoch at which
  cosmological constant started dominating over matter, so that $(a_\Lambda/a_0)^3=
  (\Omega_{DM}/\Omega_\Lambda)$. Then we find that the dynamic range of 
 the phase 3 is 
 \begin{equation}
\frac{a_F}{a_\Lambda} = 2\pi T_0 L_\Lambda \left( \frac{\Omega_\Lambda}{\Omega_{DM}}\right)^{1/3}
\approx 3\times 10^{30}
\label{twooone}
\end{equation}

 Interestingly enough, one can also impose a similar bound on the physically relevant duration of inflation. 
 We know that the quantum fluctuations, generated during this inflationary phase, could act as seeds of structure formation in the universe. Consider a perturbation at some given wavelength scale which is stretched with the expansion of the universe as $\lambda\propto a(t)$.
 During the inflationary phase, the Hubble radius remains constant while the wavelength increases, so that the perturbation will `exit' the Hubble radius at some time. In the radiation dominated phase, the Hubble radius $d_H\propto t\propto a^2$ grows faster than the wavelength $ \lambda\propto a(t)$. Hence, normally, the perturbation will `re-enter' the Hubble radius at some time.
 If there was no re-inflation,  {\it all} wavelengths will re-enter the Hubble radius sooner or later.
 But if the universe undergoes re-inflation, then the Hubble radius `flattens out' at late times and some of the perturbations will {\it never} reenter the Hubble radius ! If we use the criterion that we need the perturbation to reenter the Hubble radius, we get a natural bound on the duration of inflation which is of direct astrophysical relevance.  Consider  a perturbation which leaves the Hubble radius ($H_{in}^{-1}$) during the inflationary epoch at $a= a_i$. It will grow to the size $H_{in}^{-1}(a/a_i)$ at a later epoch. 
 We want to determine $a_i$ such that this length scale grows to 
   $L_\Lambda$ just when the dark energy starts dominating over matter; that is at
 the epoch $a=a_\Lambda = a_0(\Omega_{DM}/\Omega_{\Lambda})^{1/3}$. 
  This gives 
  $H_{in}^{-1}(a_\Lambda/a_i)=L_\Lambda$ so that $a_i=(H_{in}^{-1}/L_\Lambda)(\Omega_{DM}/\Omega_{\Lambda})^{1/3}a_0$. On the other hand, the inflation ends at 
  $a=a_{end}$ where $a_{end}/a_0 = T_0/T_{\rm reheat}$ where $T_{\rm reheat} $ is the temperature to which the universe has been reheated at the end of inflation. Using these two results we can determine the dynamic range of this phase 1 to be 
  \begin{equation}
\frac{a_{\rm end} }{a_i} = \left( \frac{T_0 L_\Lambda}{T_{\rm reheat} H_{in}^{-1}}\right)
\left( \frac{\Omega_\Lambda}{\Omega_{DM}}\right)^{1/3}=\frac{(a_F/a_\Lambda)}{2\pi T_{\rm reheat} H_{in}^{-1}} \cong 10^{25}
\label{twootwo}
\end{equation} 
where we have used the fact that, for a GUTs scale inflation with $E_{GUT}=10^{14} GeV,T_{\rm reheat}=E_{GUT},\rho_{in}=E_{GUT}^4$
we have $2\pi H^{-1}_{in}T_{\rm reheat}=(3\pi/2)^{1/2}(E_P/E_{GUT})\approx 10^5$.
If we consider a quantum gravitational, Planck scale, inflation with $2\pi H_{in}^{-1} T_{\rm reheat} = \mathcal{O} (1)$, the ranges in \eq{twooone} and \eq{twootwo} are approximately equal. 

This fact is definitely telling us something regarding the duality between Planck scale and Hubble scale or between the infrared and ultraviolet limits of the theory.
The mystery is compounded by the fact the asymptotic De Sitter phase has an observer dependent horizon and
related thermal properties \cite{ghds}. Recently, it has been shown --- in a series of papers, see ref.\cite{tpholo} ---  that it is possible to obtain 
classical relativity from purely thermodynamic considerations.  It is difficult to imagine that these features are unconnected and accidental; at the same time, it is difficult to prove a definite connection between these ideas and the \cc.

\subsection{Gravity as detector of the vacuum energy}

Finally, we  will describe an idea which \textit{does} lead to the correct value of \cc.
The conventional discussion of the relation between cosmological constant and vacuum energy density is based on
evaluating the zero point energy of quantum fields with an ultraviolet cutoff and using the result as a 
source of gravity.
Any reasonable cutoff will lead to a vacuum energy density $\rho_{\rm vac}$ which is unacceptably high. 
This argument,
however, is too simplistic since the zero point energy --- obtained by summing over the
$(1/2)\hbar \omega_k$ --- has no observable consequence in any other phenomena and can be subtracted out by redefining the Hamiltonian. The observed non trivial features of the vacuum state of QED, for example, arise from the {\it fluctuations} (or modifications) of this vacuum energy rather than the vacuum energy itself. 
This was, in fact,  known fairly early in the history of cosmological constant problem and is stressed by Zeldovich \cite{zeldo} who explicitly calculated one possible contribution to {\it fluctuations} after subtracting away the mean value.
This
suggests that we should consider   the fluctuations in the vacuum energy density in addressing the 
cosmological constant problem. 

If the vacuum probed by the gravity can readjust to take away the bulk energy density $\rho_{_{\rm UV}}\simeq L_P^{-4}$, quantum \textit{fluctuations} can generate
the observed value $\rho_{\rm DE}$. One of the simplest models \cite{tpcqglamda} which achieves this uses the fact that, in the semi-classical limit, the wave function describing the universe of proper four-volume ${\cal V}$ will vary as
$\Psi\propto \exp(-iA_0) \propto 
 \exp[ -i(\Lambda_{\rm eff}\mathcal V/ L_P^2)]$. If we treat 
  $(\Lambda/L_P^2,{\cal V})$ as conjugate variables then uncertainty principle suggests $\Delta\Lambda\approx L_P^2/\Delta{\cal V}$. If
the four volume is built out of Planck scale substructures, giving $ {\cal V}=NL_P^4$, then the Poisson fluctuations will lead to $\Delta{\cal V}\approx \sqrt{\cal V} L_P^2$ giving
    $ \Delta\Lambda=L_P^2/ \Delta{\mathcal V}\approx1/\sqrt{{\mathcal V}}\approx   H_0^2
 $. (This idea can be a more quantitative; see \cite{tpcqglamda}).

Similar viewpoint arises, more rigorously, when we study the question of \emph{detecting} the energy
density using gravitational field as a probe.
 Recall that an Unruh-DeWitt detector with a local coupling $L_I=M(\tau)\phi[x(\tau)]$ to the {\it field} $\phi$
actually responds to $\langle 0|\phi(x)\phi(y)|0\rangle$ rather than to the field itself \cite{probe}. Similarly, one can use the gravitational field as a natural ``detector" of energy momentum tensor $T_{ab}$ with the standard coupling $L=\kappa h_{ab}T^{ab}$. Such a model was analysed in detail in ref.~\cite{tptptmunu} and it was shown that the gravitational field responds to the two point function $\langle 0|T_{ab}(x)T_{cd}(y)|0\rangle $. In fact, it is essentially this fluctuations in the energy density which is computed in the inflationary models (see \eq{cnumber}) as the seed {\it source} for gravitational field, as stressed in
ref.~\cite{tplp}. All these suggest treating the energy fluctuations as the physical quantity ``detected" by gravity, when
one needs to incorporate quantum effects.  
If the \cc\ arises due to the energy density of the vacuum, then one needs to understand the structure of the quantum vacuum at cosmological scales. Quantum theory, especially the paradigm of renormalization group has taught us that the energy density --- and even the concept of the vacuum
state --- depends on the scale at which it is probed. The vacuum state which we use to study the
lattice vibrations in a solid, say, is not the same as vacuum state of the QED.

 In fact, it seems \textit{inevitable} that in a universe with two length scale $L_\Lambda,L_P$, the vacuum
 fluctuations will contribute an energy density of the correct order of magnitude $\rho_{_{\rm DE}}=\sqrt{\rho_{_{\rm IR}}\rho_{_{\rm UV}}}$. The hierarchy of energy scales in such a universe, as detected by 
 the gravitational field has \cite{plumian,tpvacfluc}
 the pattern
 \begin{equation}
\rho_{\rm vac}={\frac{1}{ L^4_P}}    
+{\frac{1}{L_P^4}\left(\frac{L_P}{L_\Lambda}\right)^2}  
+{\frac{1}{L_P^4}\left(\frac{L_P}{L_\Lambda}\right)^4}  
+  \cdots 
\end{equation}  
 The first term is the bulk energy density which needs to be renormalized away (by a process which we  do not understand at present); the third term is just the thermal energy density of the De Sitter vacuum state; what is interesting is that quantum fluctuations in the matter fields \textit{inevitably generate} the second term.

The key new ingredient arises from the fact that the properties of the vacuum state  depends on the scale at which it is probed and it is not appropriate to ask questions without specifying this scale. 
 If the spacetime has a cosmological horizon which blocks information, the natural scale is provided by the size of the horizon,  $L_\Lambda$, and we should use observables defined within the accessible region. 
The operator $H(<L_\Lambda)$, corresponding to the total energy  inside
a region bounded by a cosmological horizon, will exhibit fluctuations  $\Delta E$ since vacuum state is not an eigenstate of 
{\it this} operator. The corresponding  fluctuations in the energy density, $\Delta\rho\propto (\Delta E)/L_\Lambda^3=f(L_P,L_\Lambda)$ will now depend on both the ultraviolet cutoff  $L_P$ as well as $L_\Lambda$.  
 To obtain
 $\Delta \rho_{\rm vac} \propto \Delta E/L_\Lambda^3$ which scales as $(L_P L_\Lambda)^{-2}$
 we need to have $(\Delta E)^2\propto L_P^{-4} L_\Lambda^2$; that is, the square of the energy fluctuations
 should scale as the surface area of the bounding surface which is provided by the  cosmic horizon.  
 Remarkably enough, a rigorous calculation \cite{tpvacfluc} of the dispersion in the energy shows that
 for $L_\Lambda \gg L_P$, the final result indeed has  the scaling 
 \begin{equation}
 (\Delta E )^2 = c_1 \frac{L_\Lambda^2}{L_P^4} 
 \label{deltae}
 \end{equation}
 where the constant $c_1$ depends on the manner in which ultra violet cutoff is imposed.
 Similar calculations have been done (with a completely different motivation, in the context of 
 entanglement entropy)
 by several people and it is known that the area scaling  found in Eq.~(\ref{deltae}), proportional to $
L_\Lambda^2$, is a generic feature \cite{area}.
For a simple exponential UV-cutoff, $c_1 = (1/30\pi^2)$ but  cannot be computed
 reliably without knowing the full theory.
  We thus find that the fluctuations in the energy density of the vacuum in a sphere of radius $L_\Lambda$ 
 is given by 
 \begin{equation}
 \Delta \rho_{\rm vac}  = \frac{\Delta E}{L_\Lambda^3} \propto L_P^{-2}L_\Lambda^{-2} \propto \frac{H_\Lambda^2}{G}
 \label{final}
 \end{equation}
 The numerical coefficient will depend on $c_1$ as well as the precise nature of infrared cutoff 
 radius (like whether it is $L_\Lambda$ or $L_\Lambda/2\pi$ etc.). It would be pretentious to cook up the factors
 to obtain the observed value for dark energy density. 
 But it is a fact of life that a fluctuation of magnitude $\Delta\rho_{vac}\simeq H_\Lambda^2/G$ will exist in the
energy density inside a sphere of radius $H_\Lambda^{-1}$ if Planck length is the UV cut off. {\it One cannot get away from it.}
On the other hand, observations suggest that there is a $\rho_{vac}$ of similar magnitude in the universe. It seems 
natural to identify the two, after subtracting out the mean value by hand. Our approach explains why there is a \textit{surviving} cosmological constant which satisfies 
$\rho_{_{\rm DE}}=\sqrt{\rho_{_{\rm IR}}\rho_{_{\rm UV}}}$
 which ---  in our opinion --- is {\it the} problem.

\section*{Acknowledgement}

I thank J. Alcaniz for his friendship and warm hospitality during the
 X Special Courses at Observatorio Nacional, Rio de Janeiro, Brazil
  and for persuading
me to write  up my lecture notes. I am grateful to Gaurang Mahajan for help in generating the figures.

\end{document}